# Effect of Magnetic Anisotropy and Gradient-Induced Dzyaloshinskii-Moriya Interaction on the Formation of Magnetic Skyrmions

*Adam Erickson,[1†] Qihan Zhang,[2†] Hamed Vakili,[3†] Edward Schwartz,[3†] Suvechhya Lamichhane,[3] Chaozhong Li,[4] Boyu Li,[5] Dongsheng Song,[5] Guozhi Chai,[4] Sy-Hwang Liou,[3] Alexey A. Kovalev,[3*] Jingsheng Chen,[2,6*] Abdelghani Laraoui[1,3*]*

[1]Department of Mechanical & Materials Engineering, University of Nebraska-Lincoln, 900 N 16th Street, W342 NH, Lincoln, NE 68588, United States
[2]Department of Materials Science and Engineering, National University of Singapore, Block E2, #05-19, 5 Engineering Drive 2, Singapore 117579, Singapore
[3]Department of Physics and Astronomy and the Nebraska Center for Materials and Nanoscience, University of Nebraska-Lincoln, 855 N 16th St, Lincoln, NE 68588, United States
[4]Key Laboratory for Magnetism and Magnetic Materials of Ministry of Education, School of Physical Science and Technology, Lanzhou University, Lanzhou 730000, China
[5]Institutes of Physical Science and Information Technology, Anhui University, Hefei, 230601, China
[6]National University of Singapore (Suzhou) Research Institute, Suzhou, Jiangsu, 215123, China
[†]Equal contributions
[*]Corresponding authors: akovalev2@nebraska.edu, msecj@nus.edu.sg, alaraoui2@unl.edu

## Abstract

Topological spin textures (e.g. skyrmions) can be stabilized by interfacial Dzyaloshinskii-Moriya interaction (DMI) in the magnetic multilayer, which has been intensively studied. Recently, Bloch-type magnetic skyrmions stabilized by composition gradient-induced DMI (g-DMI) have been observed in 10-nm thick CoPt single layer. However, magnetic anisotropy in gradient-composition engineered CoPt (g-CoPt) films is highly sensitive to both the relative Co/Pt composition and the film thickness, leading to a complex interplay with g-DMI. The stability of skyrmions under the combined influence of magnetic anisotropy and g-DMI is crucial yet remains poorly understood. Here, we conduct a systematic study on the characteristics of magnetic skyrmions as a function of gradient polarity and effective gradient (defined as gradient/thickness) in g-CoPt single layers (thickness of 10 – 30 nm) using magnetic force microscopy (MFM), bulk magnetometry, and topological Hall effect measurements. Brillouin light scattering spectroscopy confirms that both the sign and magnitude of g-DMI depend on the polarity and amplitude of the composition gradient in g-CoPt films. MFM reveals that skyrmion size and density vary with g-CoPt film thickness, gradient polarity, and applied magnetic field. An increased skyrmion density is observed in samples exhibiting higher magnetic anisotropy, in agreement with micromagnetic simulations and energy barrier calculations.

## 1. Introduction

Magnetic skyrmions are topologically protected, nanoscale spin textures that have garnered considerable interest for promising potential for the next generation of ultra-dense and energy-efficient spintronic devices.[1,2] These spin configurations are stabilized by the Dzyaloshinskii–Moriya interaction (DMI), which arises in magnetic systems with strong spin–orbit coupling (SOC) and broken inversion symmetry.[3,4] Interfacial DMI was shown to stabilize Néel-type skyrmions in ultrathin ferromagnetic (FM) films interfaced with heavy metals (HM), which have been observed first at low temperatures in epitaxially grown Fe and PdFe magnetic layers on Ir,[5,6] and at later at room temperature in stacks of layers, composed of <1-nm-thick Co layers sandwiched between HM layers Ir, Pt, and W.[7–10] By varying the FM and/or HM layer compositions, small size (< 50 nm) and high density (~60 skyrmion/μm$^2$) skyrmions was achieved.[10] Pt/Co/Ta heterostructures with Pt-Co alloy interlayers featuring compositional gradients were also recently studied, finding that broken inversion symmetry of graded interfaces combines interfacial and bulk-like DMI contributions that depend on the alloy composition.[11] While skyrmion properties are highly sensitive to film thickness and interfacial quality, degradation of these factors can compromise the uniformity and reliability of skyrmion behavior. Magnetic skyrmions have also been observed in chiral B20 compounds that possess bulk DMI such as MnSi,[12,13] $Fe_{1-x}Co_xSi$,[14] and FeGe.[15,16] Bulk DMI-induced skyrmions offer potential for thermodynamic stability in bulk (thick) FM films at room temperature. However, they also come with challenges, such as the need for specific crystal structures or potential difficulties in large-scale synthesis. Furthermore, as techniques which aim to actively manipulate DMI strength through modulation of gate voltages,[17,18] strain,[19,20] spin current,[21] temperature,[22] and magnetic field orientation[23] are being developed, additional material geometries/ properties may be useful.

Recently, gradient-induced bulk DMI (g-DMI) could be stabilized by the synergistic action of SOC and composition gradient-induced bulk magnetic asymmetry (BMA) materials.[24,25] Sizeable g-DMI amplitudes were measured in gradient composition $Co_xPt_{1-x}$ (g-CoPt) single-layer systems with positive and negative composition gradient $\Delta x/t$ ($t$ is the thickness of the film), where $\Delta x$ (up to 50%) is defined as the difference in Co percentage between the bottom and top interface.[25] Such an approach opens new possibilities for engineering skyrmions with enhanced tunability and spatially varying properties. By using scanning probe based nitrogen vacancy (NV) magnetometry[26–28] and magnetic force microscopy (MFM),[29] we recently imaged Bloch-type skyrmions (size ~ 120 – 480 nm) in $\Delta x = \pm 50\%$ 10 nm g-CoPt single layers.[30] While the magnetic anisotropy in g-CoPt is highly sensitive to both the relative Co/Pt composition and the film thickness, leading to a complex relationship with g-DMI,[24,25,30] which in turn may affect the properties of magnetic skyrmions. Therefore, elucidating the effect of magnetic anisotropy and g-DMI on skyrmion size and density is crucial for advancing high-density magnetic memory applications.[1,2]

We prepared a series of g-CoPt single layers (thickness of 10 – 30 nm) with varying effective gradients (g-DMI), accompanied by corresponding changes in magnetic anisotropy. The reversal of the gradient polarity results in the opposite sign of the Dzyaloshinskii-Moriya vector, $D_{ijl}$, confirmed by Brillouin light scattering (BLS) spectroscopy measurements. To study the effect of magnetic anisotropy and g-DMI on properties of skyrmions, we conducted a systematic study on the effective gradient dependent skyrmions characteristics by using MFM, bulk magnetometry, and topological Hall effect measurements. An asymmetry in the statistical nucleation of skyrmions from the uniform and polydomain starting configurations was observed. Then, an increased skyrmion density was observed in films exhibiting higher magnetic anisotropy, consistent with micromagnetic simulations of skyrmion relaxation and energy barrier calculations. These results reveal that dipolar interactions have an important role in the stabilization of skyrmions in thicker g-CoPt films and g-DMI primarily determines their helicity.

## 2. Results and Discussion

### 2.1. Structural and magnetic characterization of g-CoPt films

From previous studies,[24,25,30,31] there is a proportionality between the effective gradient and the resulting g-DMI. It is noted that the effective gradient is defined as composition gradient of Co ($\Delta x$)/thickness ($t$). Thus, for a fixed value of $\Delta x$, the g-DMI can be modulated by changing $t$. In this study, $SiO_2$ (2 nm)/$Co_xPt_{1-x}$ (thickness $t$ = 10, 20, 30 nm) with fixed gradient parameters of opposite sign ($\Delta x = \pm 50\%$) were deposited on a $SrTiO_3$ (STO) (111) single-crystal substrate (see Experimental Section for the growth details).[30] Considering the hcp structure, the lattice constants are 5.19 Å, 5.472 Å and 5.529 Å for $Co_3Pt$, $CoPt_3$, and STO substrate respectively. Due to the small lattice mismatch between $Co_xPt_{1-x}$ and STO, STO substrate is selected for the preparation of our g-CoPt films. The used composition difference $\Delta x$ of +50% (-50%) corresponds to $CoPt_3 \rightarrow Co_3Pt$ ($Co_3Pt \rightarrow CoPt_3$) from the start to end of the growth, see Figure 1a. Figure 1b shows X-ray diffraction (XRD) spectra of the $\Delta x = \pm 50\%$ 20 and 30 nm thick g-CoPt films. The XRD spectra of the $\Delta x = \pm 50\%$ 10 nm g-CoPt films were reported in our previous work.[30] There is significant broadening compared to binary (non-gradient) CoPt films, which may be affected by nonuniform composition or a lack of long-range crystallinity.[32,33] Atomic force microscopy (AFM) topography measurements on $\Delta x = \pm 50\%$ 20 and 30 nm g-CoPt films revealed smooth surfaces with a roughness in the range of 0.48 – 1.48 nm (see Supporting Information Section S1 and Figure S1.1).

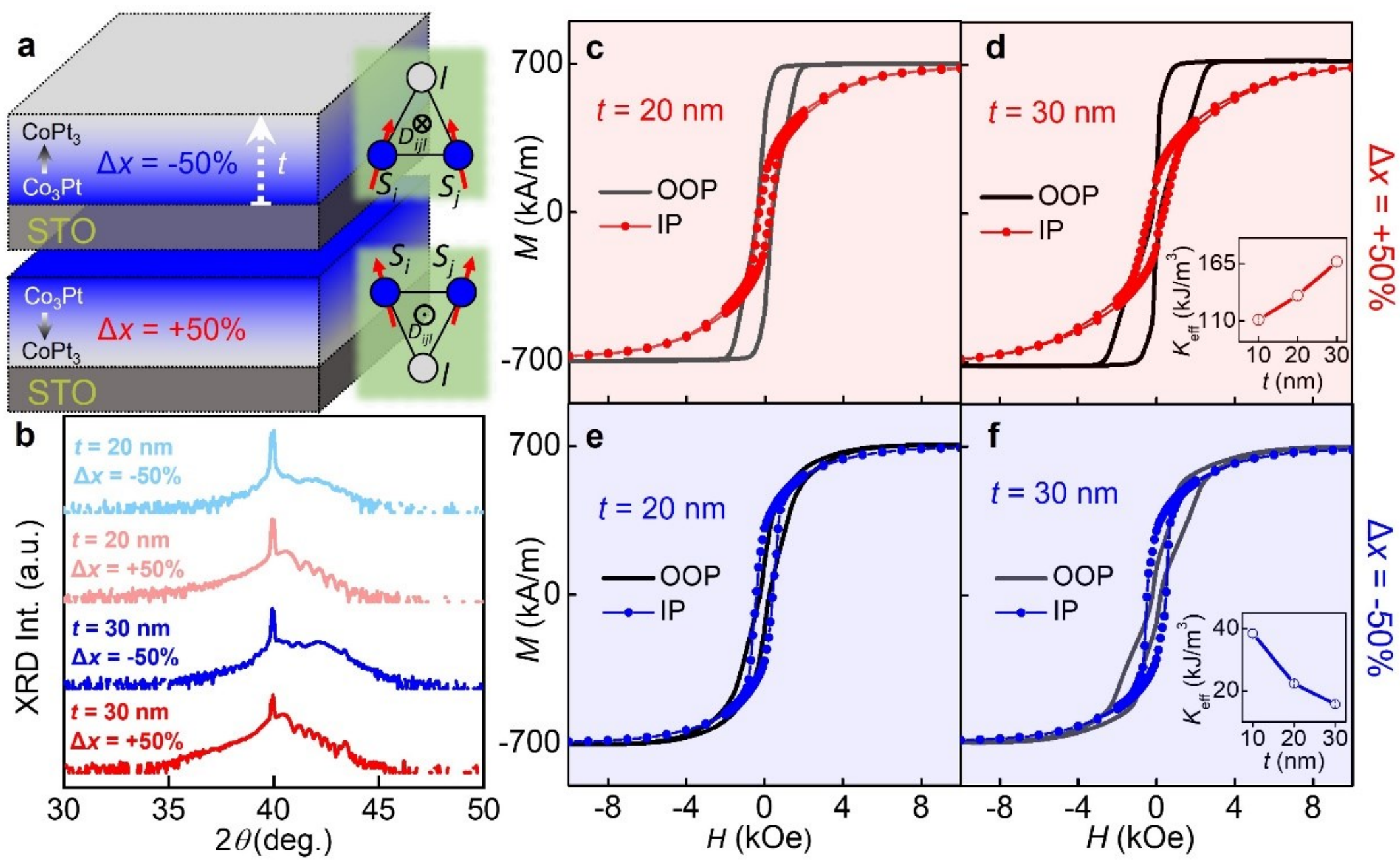

**Figure 1. Structural and magnetic characterization of g-CoPt films**. (a) Cartoon depiction of the sample geometry, indicating the defined gradient parameter, $\Delta x$, for negatively and positively graded CoPt films. The statistically averaged asymmetric exchange sites between spins ($S_i$,$S_j$) and heavy metal site ($l$) result in a non-zero DMI term ($D_{ijl}$) whose sign is dependent on the polarity of the gradient. (b) XRD spectra taken on 20 and 30 nm g-CoPt films with $\Delta x \pm 50\%$. IP and OOP magnetization hysteresis loops of $\Delta x = +50\%$ 20 nm (c) and 30 nm (d) g-CoPt films. The corresponding $\Delta x$ = -50% films' IP and OOP hysteresis loops are shown in e-f. Effective magnetic anisotropy $K_{eff}$ vs thickness for $\Delta x = +50\%$ (inset of d) and $\Delta x$ = -50% (inset of f) g-CoPt single layers.

Energy dispersive X-ray spectroscopy (EDS) in a scanning transmission electron microscopy (STEM) configuration was performed on selected $\Delta x$ = +50% g-CoPt films (thickness of 20 and 30 nm) to confirm the composition gradient with a relative ratio of Co to Pt of 3:1, see Supporting Information Section S1 and Figure S1.2.

The out of plane (OOP) and in-plane (IP) *M*-*H* hysteresis loops were performed using Superconducting Quantum Interference Device (SQUID) magnetometry and displayed in Figure 1c (Figure 1e) and Figure 1d (Figure 1f) for $\Delta x$ = +50% ($\Delta x$ = -50%) 20 and 30 nm g-CoPt films, respectively. The effective magnetic anisotropy, $K_{\text{eff}} = M_s \times \Delta/\mu_0$, is determined by taking the difference of area under OOP and IP magnetic hysteresis loops in Figure 1(c-f).[30,34] $M_s$ is the saturation magnetization ~700 kA/m for all $\Delta x$ = ±50% g-CoPt films. As shown in the insets of Figure 1d and Figure 1f, $K_{eff}$ = 111 ± 1, 134.5 ± 4, and 167 ± 3 kJ/m$^3$ for 10, 20, and 30 nm $\Delta x$ = +50% g-CoPt films, respectively. $K_{ef}$ = 38.4 ± 1.3, 22.3 ± 0.3, and 15.7 ± 0.47 kJ/m$^3$ for 10, 20, and 30 nm $\Delta x$ = -50% g-CoPt films, respectively. Since the bulk perpendicular magnetic anisotropy (PMA) of CoPt film relies on the close-packed stackings of Co/Pt, the increase of hexagonal close packed (HCP) phase order degree gives rise to the higher $K_{eff}$. The stacking order thus influences crystallographic quality due to the lattice mismatch among $Co_3Pt$, $CoPt_3$, and the STO substrate. If $CoPt_3$ layer is initially grown, it could be regarded as the seed layer to promote the subsequent growth of HCP phase of $Co_xPt_{1-x}$ alloy.[35–37] It results in that $K_{eff}$ of $\Delta x$ = +50% ($CoPt_3 \rightarrow Co_3Pt$) films is higher than $\Delta x$ = -50% ($Co_3Pt \rightarrow CoPt_3$) films (see the insets of Fig.1d and Fig. 1f). The relatively lower crystallinity for the $\Delta x$ = -50% ($Co_3Pt \rightarrow CoPt_3$) films agrees with the XRD spectra in Figure 1b. Furthermore, increasing the film thickness promotes (suppresses) the HCP phase formation in the $\Delta x = +50\%$ ($-50\%$) films.

BLS spectroscopy was used to estimate the value of g-DMI in all $\Delta x$ = ±50% 10 – 30 nm g-CoPt films. The nonreciprocal frequency shift, $\Delta f$, between Stokes and anti-Stokes peaks in the BLS spectra is related to the DMI energy density as:[25,30] $\Delta f = 2\gamma D k_x/\pi M_s$, where $\gamma$, $D$, and $k_x$ are the gyromagnetic ratio, the volume-averaged DMI constant, and the projection of the spin-wave vector ($\boldsymbol{k}$) in the direction perpendicular to the applied magnetic field $\mu_0 H$, respectively.[30] The representative BLS spectra for negative (positive) gradient 20 nm films are shown in Figures 2a(b). $\Delta f$ was measured as function of $k_x$ and plotted in Figure 2c for a selected g-CoPt films, in which the applied magnetic field is larger than magnetic anisotropy field. Here, the prominent aspects of g-DMI are in accordance with the previous experimental and theoretical determinations, namely that the sign of $\Delta f$, and thus the sign of $D$, is negative for a negative gradient parameter. The magnitude of g-DMI $D$ is plotted versus the effective gradient parameter $\Delta x/t$ in Figure 2d.

As expected, $D$ increases with increasing $\Delta x/t$, from -0.16±0.10 mJ/m$^2$ for the negative gradient 20 nm film to -0.51±0.06 mJ/m$^2$ for the 10 nm film. For the positive gradient samples, $D$ increases from 0.04±0.08 mJ/m$^2$ for the 30 nm film to ~ 0.15±0.05 mJ/m$^2$ for the 10 nm film. The high deviation of DMI values between the positive and negative gradient films may be explained by additional long-range structural asymmetries along the thickness direction in the film, such as crystal phase, and magnetic anisotropy, which compromises the coherence and lifetime of spin waves and consequently reduces the BLS signal strength while increasing the noise. For the 30 nm films, a measurable value of $\Delta f$ could only be obtained from $\Delta x$ = +50% g-CoPt (see Supporting Information Section S2), showing a magnitude of $D$ of 0.04±0.08 mJ/m$^2$, almost comparable with ungraded films (D = -0.055±0.041 mJ/m$^2$ for $Co_3Pt$ and = 0.008±0.001 mJ/m$^2$ for $CoPt_3$).[30] Bloch-type skyrmions can be stabilized in the 10 nm g-film films by g-DMI,[30] however for thicker g-CoPt films (e.g., $\Delta x$ = +50% 30 nm), skyrmions can be still stabilized even with close to zero g-DMI. We note the absence of magnetic skyrmions in the non-graded $CoPt_3$ and $Co_3Pt$, suggesting that g-DMI is needed to stabilize them.[30]

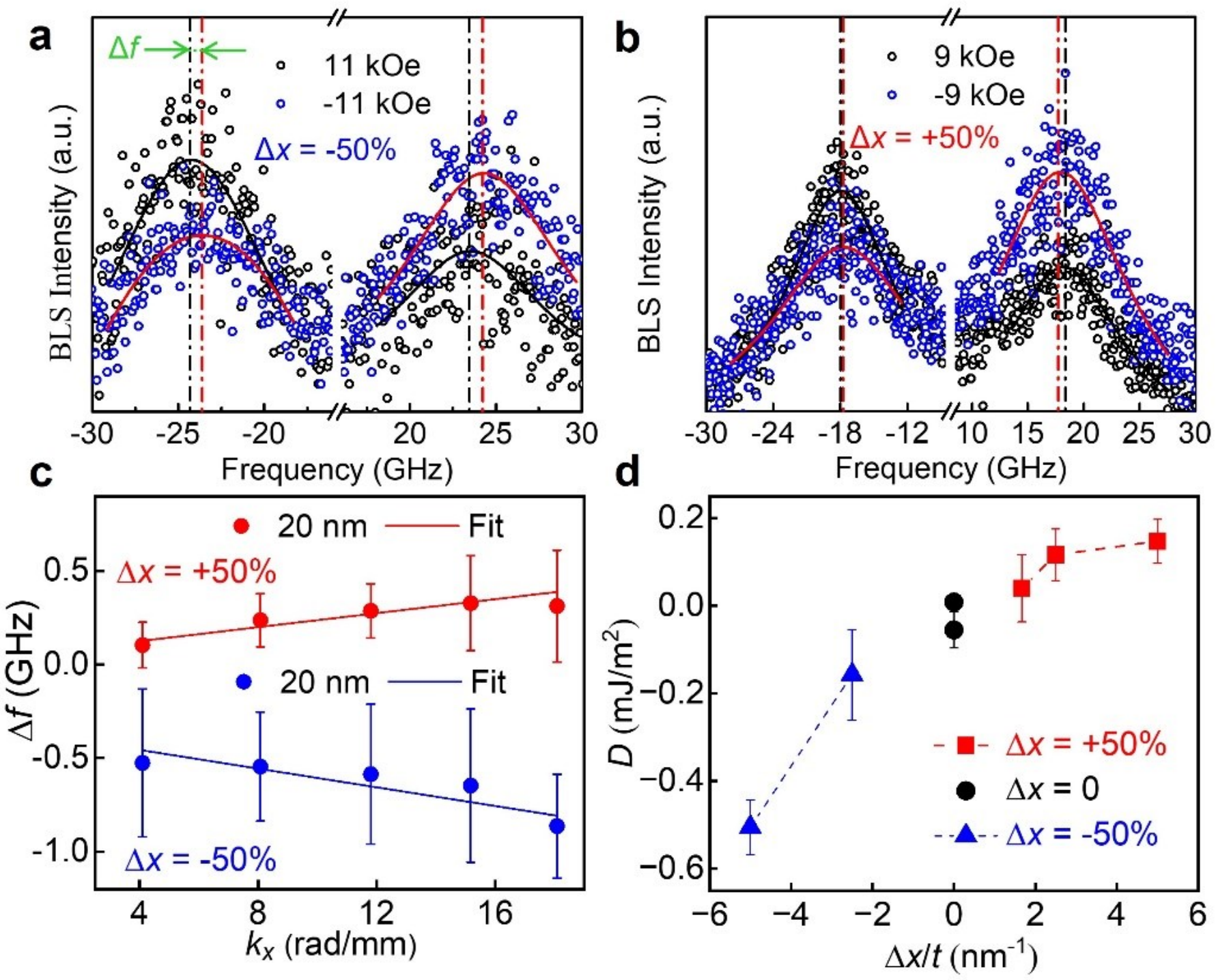

**Figure 2.** Summary of g-DMI measurements on g-CoPt films of different thicknesses: BLS spectra for $\Delta x = -50\%$ (a) and $\Delta x = +50\%$ (b) 20 nm g-CoPt film in which the scattered black and blue curves represent the spectra with $\pm\ \mu_0 H$ and the solid lines are the fitting curves. (c) $\Delta f$ vs $k_x$ for $\Delta x = \pm\ 50\%$ 20 nm g-CoPt films. The sign and magnitude of $\Delta f/k_x$ is reflected in the calculated g-DMI energy. (d) $D$ vs the effective gradient parameter, $\Delta x/t$, in all graded (10 – 30 nm) and non-graded CoPt films.

## 2.2. The evolution of spin textures via MFM imaging

We next investigated the evolution of topological spin textures by independent magneto transport (Physical Property Measurement System, PPMS) and magneto-optical Kerr effect (MOKE) measurements. The interaction between skyrmions and charge carriers gives rise to the topological Hall effect (THE), which is reflected in the residual resistivity $\Delta\rho_{xy}(H) = \rho_{xy}(H) - [R_0 H + R_s M(H)]$, where $\rho_{xy}(H)$ is the total Hall resistivity, and $R_0 H$ and $R_s M(H)$ are the ordinary and anomalous Hall contributions, respectively. The residual resistivity $\Delta\rho_{xy}(H)$ was estimated by subtracting the fitted background $\rho_{xy}^{fit}(H) = R_0 H + R_s M(H)$ from the measured $\rho_{xy}(H)$. Here $\rho_{xy}(H)$ was obtained from PPMS transport measurements, while $M(H)$ was derived from MOKE measurements (see Experimental Section and References [30,38] for further details). Since the transport and MOKE measurements were performed using different setups, we compensated for the mismatch in magnetic field calibration using a scaling factor of 0.85. See Supporting Information Section 3 and Figure S3.1 for further details. The normalized $\rho_{xy}(H)$, $\rho_{xy}^{fit}(H)$, and $\Delta\rho_{xy}(H)$ for 20 and 30 nm g-CoPt films are plotted in Figure 3, with the constituent measurements plotted for reference. It is worth noting that the plotted curves only represent one magnetic field sweep direction, which is denoted by the arrow in Figure 3a. From the plots of $\Delta\rho_{xy}(H)$, there appears to be non-zero signal for each film. These signals may not arise solely from THE. Contributions from polydomain states and shape anisotropy introduced by the Hall bar geometry may also be involved. There is also an asymmetry of the magnitude of $\Delta\rho_{xy}(H)$ peaks forming from uniform magnetization (decreasing magnetic field) as compared to from the polydomain state (increasing magnetic field).[30] The evolution of topological spin textures inferred from the $\Delta\rho_{xy}(H)$ signal is further supported by MFM imaging.[38]

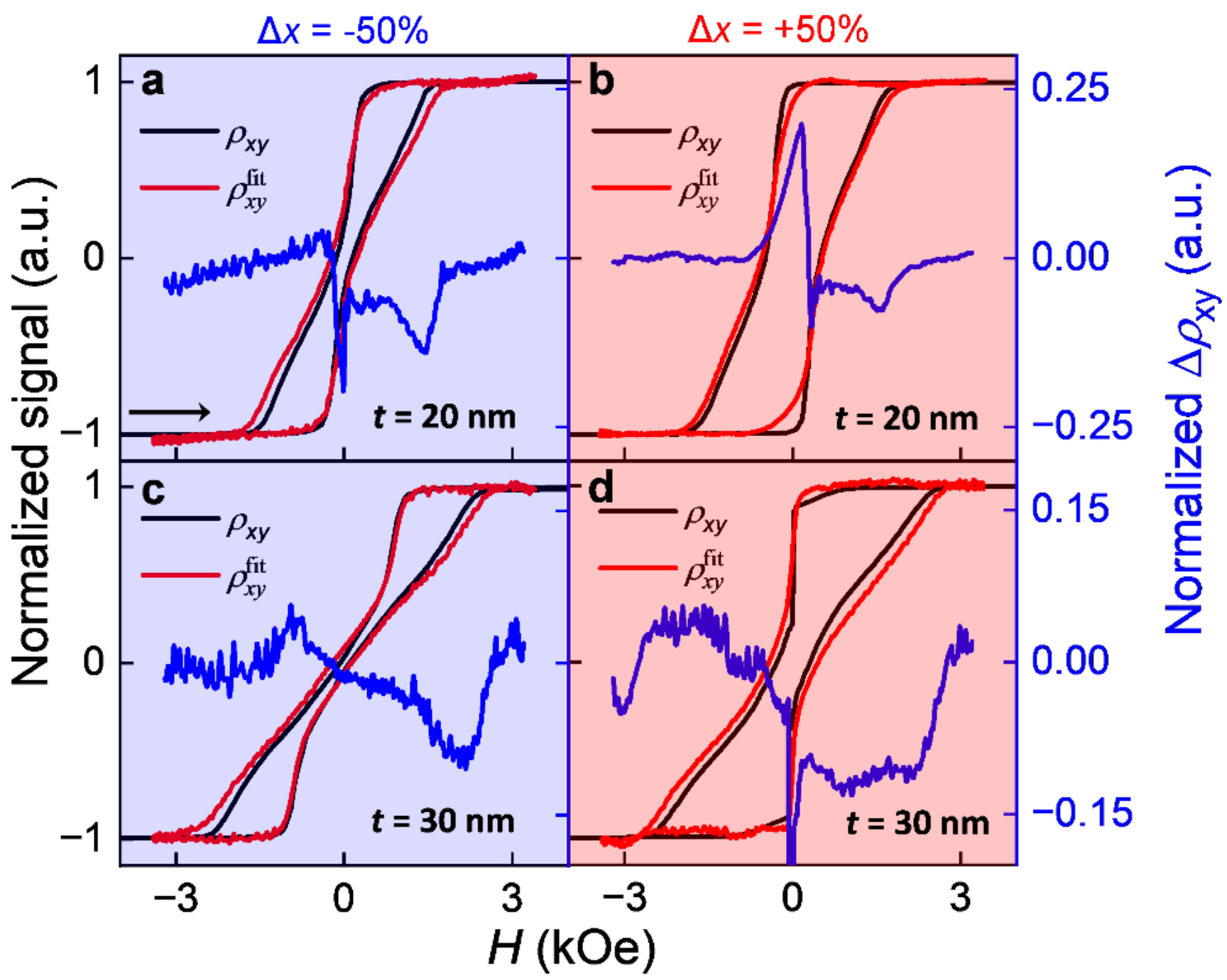

**Figure 3.** Extraction of the residual resistivity $\Delta\rho_{xy}$ from Hall effect and MOKE measurements. In each plot, $\Delta\rho_{xy}(H)$ (blue line) includes the THE signal which is obtained by subtracting $\rho_{xy}(H)$ measured by PPMS (red line) from $\rho_{xy}^{fit}(H)$ fitted from MOKE measurements (black line) for one direction of the applied magnetic field sweep (indicated by arrow in a). (a,b) Extracted $\Delta\rho_{xy}(H)$ signal for 20 nm g-CoPt films with a composition gradient of $\Delta x$ = -50% and $\Delta x$ = +50%, respectively. (c,d) Extracted $\Delta\rho_{xy}(H)$ for 30 nm g-CoPt films of $\Delta x$ = -50% and $\Delta x$ = +50%, respectively.

To obtain a clearer understanding of the underlying magnetization patterns which can give rise to THE signals, we performed MFM imaging of the magnetization reversal process over the same field sweep conditions as were used to obtain $\Delta\rho_{xy}(H)$ curves. The results for $\Delta x$ = +50% g-CoPt films ($t$ = 20, 30 nm) are shown in Figure 4. Again, it is worth noting the direction of the magnetic field sweep, which is shown in the $\Delta\rho_{xy}(H)$ plots as the blue arrows. Upon reduction of the amplitude of applied magnetic field $H$ in the uniform state, small, isolated spin textures emerge which may or may not possess a topological charge. From the MFM images, it is not possible to characterize the topology of such bubbles, and instead the density is estimated by indiscriminately counting isolated spin textures. Applying the product of relative topological charge, $Q$, the topological charge density $Q{\cdot}n_{Sk}$ can be roughly compared with the $\Delta\rho_{xy}(H)$ curves. Taking Figure 4a for 20 nm g-CoPt film , the amplitude and horizontal axis position of the roughly estimated topological density, there is some overlap with the $\Delta\rho_{xy}(H)$ curves, with the asymmetric behavior depending on the initial condition being captured as well. The asymmetry is even more pronounced in the 30 nm g-CoPt film case, where the density of skyrmions nucleated from the polydomain phase is around one order of magnitude higher than that of those nucleating from the uniform state. The comparison of MFM and $\Delta\rho_{xy}(H)$ curves is also shown for the $\Delta x$ = -50% 20, 30 nm g-CoPt films in Supporting Information Section S3 and Figure S3.2. Interestingly, the 30 nm $\Delta x$ = -50% g-CoPt film shows a distinctly low density of isolated skyrmions, though still has a prominent THE-like peak, which may stem from alternative mechanisms for topological Hall-like signals including chiral domain walls, Berry curvature, multiple hysteresis loops, or pinning effects.[39] Despite the qualitative agreement between THE and skyrmion density in positive graded films, there is no a direct prediction of the skyrmions density according to the emergent field model in the strong coupling regime, where the extracted THE resistivity component can be directly related to the density of topological spin objects.[40] This is likely due to the role of spin polarized carrier density in magneto-transport measurements,[41] as well as the residual resistivity $\Delta\rho_{xy}$ not solely related to THE (includes contributions from pinning effects induced by the edges of the Hall bar pattern) which render the simplified model insufficient. The lack of isolated skyrmions was also observed on $\Delta x$ = -50% 30 nm g-CoPt film grown on a $Al_2O_3$ substrate (see Figure S3.3 in Supporting Information). It is likely that the formation of skyrmions in these films is energetically unfavorable due to the lack of PMA. In this case, the energy barrier is likely too small to support skyrmions at room temperature and the system will favor the formation of the labyrinth domains.[42]

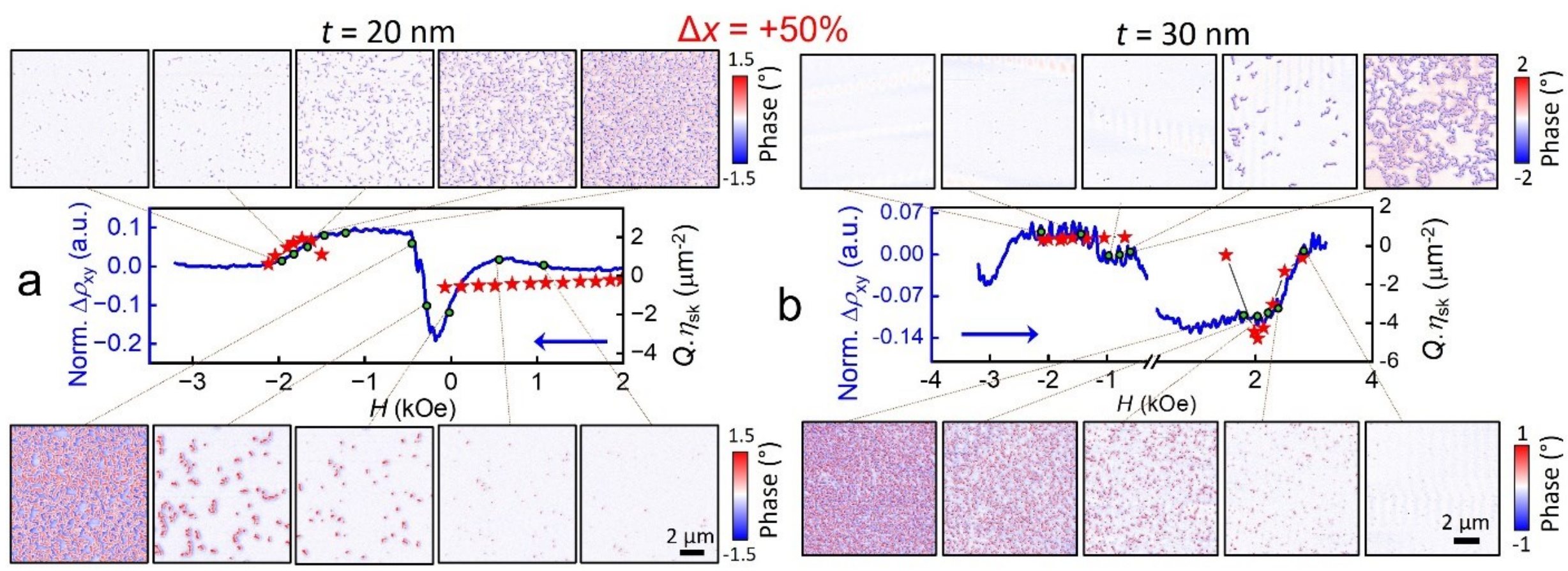

**Figure 4.** Correlation of MFM imaging and THE residual. (a) Comparison of MFM and THE measurements on $\Delta x$ = +50% 20 nm g-CoPt film. The direction of the magnetic field sweeps for both imaging and extracted THE are indicated by the blue arrow. The estimated topological charge density, $Q \cdot n_{Sk}$ is plotted as red stars. The green dots show the magnetic field at which the corresponding MFM images were taken. (b) Comparison of MFM and THE measurements on 30 nm, $\Delta x$ = +50% g-CoPt film. Note that the direction of field sweep is reversed from (a).

## 2.3 Micromagnetic Simulations and Energy Barrier Calculations

We used micromagnetic simulations to gain an additional insight of the role of film thickness on the stability, helicity, and size of spin textures in g-CoPt films using the GPU-based platform Mumax3.[43] In the simulation the difference between positive and negative gradient cases comes from the difference in magnetic anisotropy. The uniaxial anisotropy density ($K_u$) used in the simulations is related by $K_u = K_{eff} + \mu_0 M_s^2/2$ to the effective magnetic anisotropy $K_{eff}$ measured in experiments (discussed above), and magnetic dipolar interactions were switched on in simulations. In the micromagnetic simulations we used the following parameters: Exchange stiffness $A_{ex}$ = 10 J/m and mesh size is taken to be 2.5 nm × 2.5 nm × 2.5 nm. The simulations are done at zero temperature. For the minimum energy path (MEP) calculations,[44] Gilbert damping $\alpha = 1$ is used to control the energy minimization at each step.

From magnetic simulations, we find that the main effect of increasing the thickness of the film, especially in the case of $\Delta x$ = +50% g-CoPt films, is to increase the relative contribution of magnetic dipole-dipole interactions. In addition, g-CoPt films with different thicknesses exhibited variations in magnetic anisotropy (discussed in Section 2.1). Both of the above have implications for the magnetic field range in which skyrmions stabilize, as well as for their energy barrier to decay into either the uniform or maze domain phase. To account for the effect of magnetic anisotropy, we performed micromagnetic simulations for a range of values in the vicinity of the measured magnetic anisotropy values. In Figure 5a, the estimated skyrmions radius from MFM images (Figure 4 and Figure S3.2) is compared with the radius of skyrmions from simulations as a function of external magnetic field for $\Delta x$ = +50%g-CoPt films. The simulations qualitatively agree with the experimental findings, exhibiting a threshold field above which skyrmions collapse, as evidenced in Figure 5b.

Furthermore, the micromagnetic simulations presented in the Supporting Information Section S4 show that at lower magnetic fields, the energetically preferred maze state takes over, which is consistent with Figure 5b, showing the presence of skyrmions within a specific range of magnetic fields,[45,46] see for example Figure S4.3. We observe some discrepancy between the experimentally determined range of magnetic fields in which skyrmions are stable and the micromagnetic simulations, as evidenced in Figure 5b, particularly for the $\Delta x$ = +50% 20 nm g-CoPt film. These discrepancies are likely related to nonuniformities in the graded film, as well as to the presence of pinning centers, which create higher energy barriers to annihilation[47] which can affect the stability of magnetic skyrmions. Other differences may be accounted for by the fact that simulations were performed without the influence of temperature.[6] Further details on the effect of g-DMI amplitude on the radius and helicity of skyrmions, as well as plots of skyrmion radius versus $H$ for different anisotropy values, are provided in Supporting Information Section S4 and Figures S4.1, S4.2, and S4.3. The estimated density of skyrmions when sweeping from the labyrinth phase to saturation in g-CoPt films of 10, 20, 30 nm thicknesses are shown in Figure 5b, deduced from MFM images in Figure 4, Figure S3.2, and reference [30] (for 10 nm films). For $\Delta x$ = +50% g-CoPt films the combined effect of increased dipolar interactions and magnetic anisotropy results in higher skyrmion densities, as previously discussed in the context of the comparison to THE measurements. It is known that magnetic dipolar interactions can favor the formation of skyrmions.[42] Since other magnetic textures exist in the experiment as well, individual skyrmions can be affected by other magnetic textures which can further increase their stability, as evidenced in simulations. Unlike g-DMI, the lack of a well-defined handedness of dipolar interactions can give rise to a wider variety of spin textures including higher order skyrmions and antiskyrmions.[30]

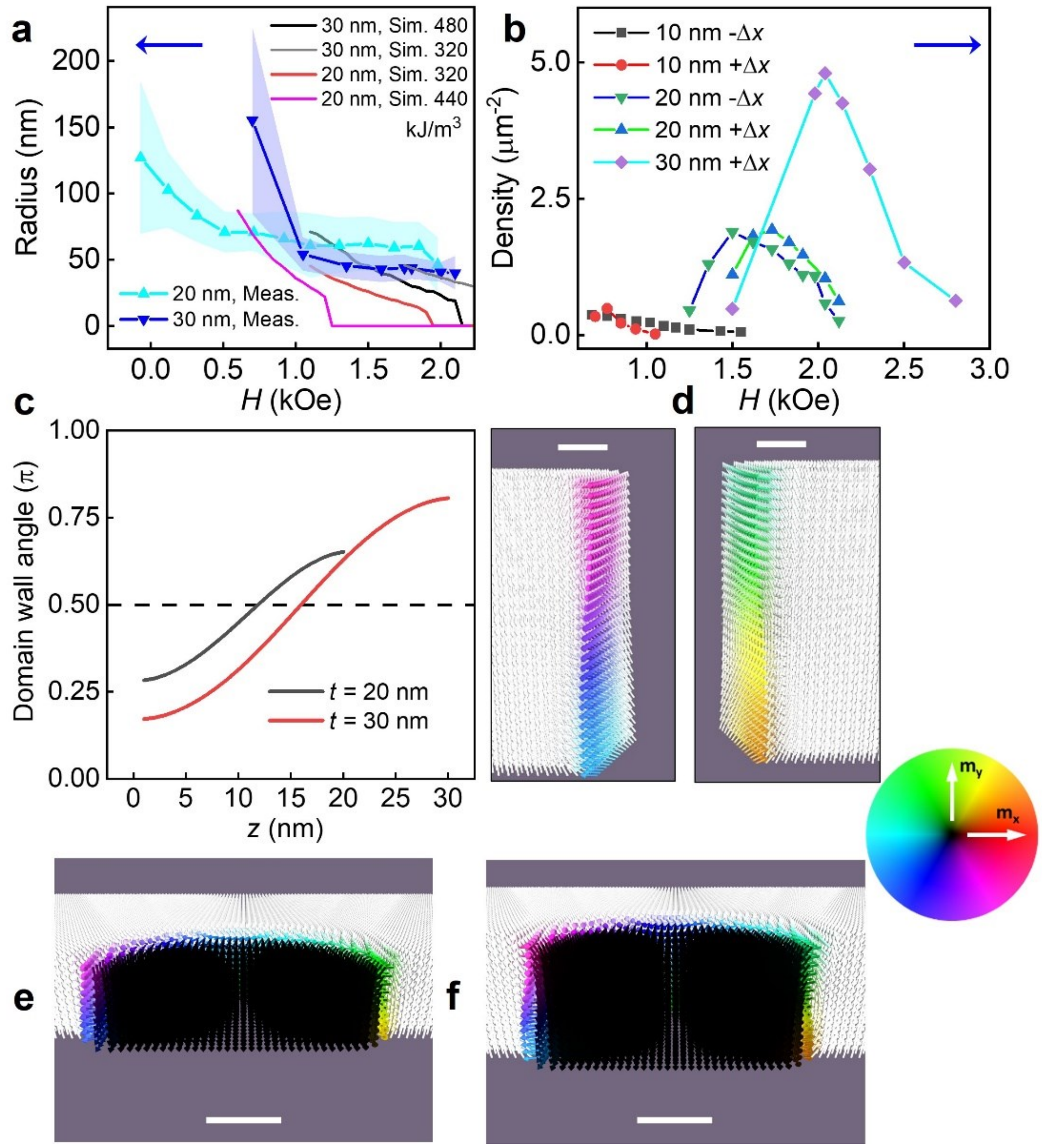

**Figure 5.** Summary of MFM and micromagnetic simulations of skyrmions in g-DMI films. (a) Measured (blue and aqua lines) and simulated (purple and black lines) skyrmion radius vs $H$ for $\Delta x$ = +50% g-CoPt of 20, 30 nm thickness. Red and grey lines are with $K_u$ at the shape demagnetization limit for 20 nm and 30 nm thickness ($K_u$ = 320 kJ/m$^3$). The experimental data is shown with a shaded area which represents the interquartile range of the measured radii. (b) Density of skyrmions from MFM imaging for 10, 20, 30 nm thick g-CoPt. Blue arrows in (a) and (b) indicate the magnetic field sweep direction. (c) Simulated twisting of the domain wall angle through the thickness for $\Delta x$ = +50% g-CoPt of 20 nm thickness with $K_u$ = 440 kJ/m$^3$, DMI = 0.12 mJ/m$^2$ and 30 nm thickness with $K_u$ = 480 kJ/m$^3$, g-DMI = 0.05 mJ/m$^2$. (d) Vector representation of cross-section of the helical skyrmion domain walls. Cross-section view of skyrmion for 20 nm (e) and 30 nm (f) $\Delta x$ = +50% g-CoPt films. The white arrows are the uniform background, and the black arrows are the core of the skyrmion. The scale bar in (d) is 5 nm and in (e-f) is 10 nm.

In the Supporting Information Section S5, we analyze the likelihood of the appearance of other textures by calculating the MEP using the string method,[44] see for example Figure S5.1, demonstrating substantial preference for the formation of skyrmions over trivial bubbles. We therefore expect that most of the observed textures in our MFM images are skyrmions as evidenced from FigureS4.4 in Supporting Information. Importantly, the densities shown are those that emerge when the magnetic field is increased from the polydomain state. The micromagnetic simulations in Figures S4.2, S4.3, and S4.4, as well as the MEP calculations in Figure S5.2, suggest that the energy cost for skyrmion formation from the polydomain state differs from that of the transition from the uniform magnetization state. In particular, when we start from the maze state, the energy cost for forming a skyrmion is close to zero at a certain magnetic field. When we start from the uniform state, there is always a finite energy cost associated with skyrmion formation. These differences in energetics may be correlated with the resulting skyrmion density, i.e., overcoming larger energy barrier may be less probable. This correlates well with the MFM results in Figure 4, where a higher density of skyrmions is observed when sweeping from the polydomain state.

For thicker simulated g-CoPt single layer, an increasingly significant rotation of the domain wall angle from the bottom to the top interface was also observed in micromagnetic simulations. This is shown in Figure 5c, where the domain wall angle is plotted versus thickness in units of π. The dashed line indicated where the helicity is Bloch-type, and the plot bounds correspond to the Néel-type. Thus, thicker films approach Néel-type helicity at the interfaces due to the minimization of stray field energy. The simulated vectorial representations of cross sections of skyrmion domain wall magnetization are shown in Figure 5d for 30 nm $\Delta x = +50\%$. Such skyrmions with twists have been observed experimentally.[48] Figure 5e (5f) shows the cross section of skyrmion for 20 nm (30 nm) $\Delta x = +50\%$ g-CoPt film and illustrates the helicity twisting. The twisted helicity in thicker films could have implications for skyrmion mobility in response to applied current density.

## 3. Conclusion

In closing, we studied a series of g-CoPt single layers of different thicknesses (10 – 30 nm) and characterized the stability of room temperature skyrmions. Firstly, we find that the $\Delta x = +50\%$ g-CoPt films possess THE residuals which correlate with the density of skyrmions observed from MFM imaging, though there is a discrepancy between these measurements for the case of $\Delta x = +50\%$ 30 nm g-CoPt film. The asymmetry of skyrmion nucleation density is also reflected in the measurements as a function of the magnetic field-sweep direction, which may be related to the difference in energy barrier between the skyrmion phase and the starting spin configuration. We also observed that the magnetic dipolar interactions in thick g-CoPt films can drive the formation of a higher density of skyrmions even though g-DMI is reduced with increasing the film thickness.

The imperfect alignment of the topological Hall effect contribution and the measured skyrmion density is attributed to sample inhomogeneity and additional topological Hall signatures including domain walls. These results aim to provide clarity regarding the stability of topological spin textures in samples with g-DMI and may help to inform the appropriate applications of skyrmions in compositional gradient films and directions for subsequent experiments. Examples of such follow up experiments include monitoring the skyrmion Hall angle upon injection of current[49] as well as investigation of the extreme cases of effective gradient, where g-DMI is either maximized or negligible and can be used to control the helicity of skyrmions.[25,30,50]

## 4. Experimental Section

Compositional gradient engineered CoPt, g-CoPt, (thickness $t = 10$, 20, and 30 nm) films were grown on STO (111) substrates by using d.c. and radio-frequency magnetron sputtering (Kurt J.

Lesker). The relative deposition rates of the Co and Pt elements were linearly changed during the growth, leading to a linear composition-magnetization change along the growth direction.[30] During the growth, the temperature and the Ar gas pressure were kept constant at 280 °C and 6 mTorr, respectively. A 2-nm $SiO_2$ capping layer was then deposited by radio-frequency magnetron sputtering to prevent any oxidation effect after the g-CoPt films were cooled down to room temperature. Magnetic properties were measured by a SQUID (Quantum design MPMS3). X-ray diffraction measurements were performed at room temperature at the Singapore Synchrotron Light Source with an X-ray wavelength of 1.541 Å. The CoPt films were patterned into a Hall bar with a width of 10 μm by using an Ultraviolet Maskless Lithography machine (TuoTuo Technology) and the ion beam etching technology.

The topological hall effect (THE) signal $\rho_{TH}(H)$ was extracted based on the relation: $\rho_{xy}(H) = R_0H + R_sM(H) + \rho_{TH}(H)$, where $R_0H$ and $R_sM(H)$ are the ordinary and anomalous Hall components, respectively, and $\rho_{TH}(H)$ is the resistivity from the contribution of THE. $\rho_{xy}(H)$ was measured using a Physical Property Measurement System (PPMS) with a DC current of 200 μA, under a swept out-of-plane magnetic field $H$. The residual resistivity $\Delta\rho_{xy}(H)$ was estimated by subtracting the fitted background $\rho_{xy}^{fit}(H) = R_0H + R_sM(H)$ from the experimental data, which included the contribution of THE signal $\rho_{TH}(H)$. The magnetic hysteresis loops $M(H)$ for g-CoPt single layers were measured using MOKE at the center of the Hall bar.

**Acknowledgements:** A.L., A.K., and S.-H.L. acknowledge support by the National Science Foundation EPSCoR RII Track-1: Emergent Quantum Materials and Technologies (EQUATE), Award OIA-2044049. A.L. acknowledges additional support from NSF Award 2328822. The research done at University of Nebraska-Lincoln was performed in part in the Nebraska Nanoscale Facility: National Nanotechnology Coordinated Infrastructure and the Nebraska Center for Materials and Nanoscience (and/or NERCF), supported by the National Science Foundation ECCS under Award 2025298, and the UNL Grand Challenges. The research performed in the National University of Singapore was supported by the Singapore Ministry of Education MOE-T2EP50223-0006, MOE-T2EP50123-0012, MOE-T2EP50121-0011, and MOE Tier 1: 22-4888-A0001.

**Conflict of Interest:** The authors declare no conflict of interest.

**Data Availability Statement:** The data that support the findings of this study are available from the corresponding author upon reasonable request.

**Author Contributions:** A.E. performed MFM measurements and analyzed the data with assistance of S.L. and S.-H.L.; Q.Z, grew the g-CoPt films and performed topography (AFM), XRD, MOKE, SQUID, AHE, and THE measurements; H.V., E.S., and A.K. performed micromagnetic modeling and skyrmion relaxation/energy barrier calculations; C.L. and G.C. performed BLS measurements; B.L. and D.S. performed HAADF-STEM and EDS on the graded CoPt films; A.L., J.C., and A.K. designed the experiments and supervised the project. A.E. and A.L. wrote the paper with contributions and feedback from all authors.

# Supporting Information

## S1.1 Topography of gradient-composition engineered CoPt (g-CoPt ) films

Atomic force microscope (AFM) images acquired from $\Delta x = \pm 50\%$ 20 and 30 nm thick g-CoPt films grown on $SrTiO_3$ (STO) substrate are shown in Figure S1.1 with the root mean square surface roughness, $R_q$, of each image, indicated in the inset. While the 20 nm, $\Delta x = +50\%$ film showed a larger (~1.48 nm) surface roughness, the magnetic textures did not exhibit any correlation to topography. The other films displayed sub-nanometer roughness, less smoother than $\Delta x = \pm 50\%$ 10 nm g-CoPt films (Ra ~110 – 234 pm) studied in reference [30].

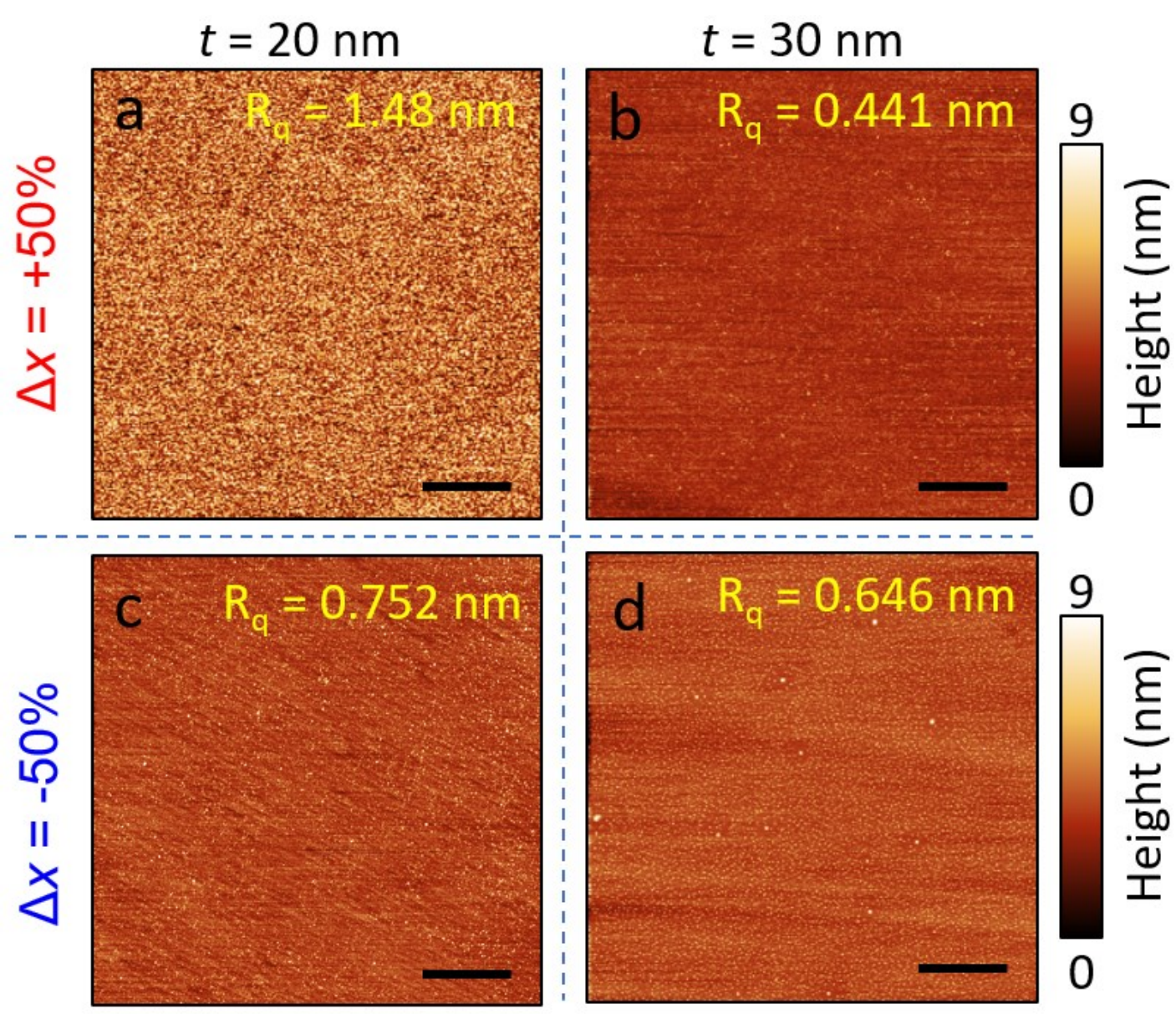

**Figure S1.1:** AFM images of 20 nm (a) and 30 nm (b) $\Delta x = +50\%$ g-CoPt films. AFM images of 20 nm (c) and 30 nm (d) $\Delta x = -50\%$ g-CoPt. The scale bar in all images is 2 μm.

## S1.2 Structural characterization of g-CoPt films

To map the composition gradient of the g-CoPt films, we performed energy dispersive X-ray spectroscopy (EDS) in a scanning transmission electron microscopy (STEM) configuration on selected $\Delta x = +50\%$ g-CoPt films (thickness of 20 nm and 30 nm). The sample specimens were prepared by Ga focused ion beam (FIB) milling. The EDS spectra are obtained on a probe/image aberration-corrected scanning transmission electron microscope (STEM). The EDS maps shown in Figure S1.2a for 20 nm film and Figure S1.2c for 30 nm film confirm the presence of Co and Pt gradient through the film thickness. The normalized EDS intensity, integrated across the areas in Figure S1.2b (for the 20 nm film) and Figure S1.2b (for the 30 nm film), show the evolution of the composition of Co and Pt through the thickness of the g-CoPt films with a relative ratio of Co to Pt is approximately 3:1, which agrees with theoretical values and with those measured in $\Delta x = \pm 50\%$ 10 nm g-CoPt films.[30] The compositional gradient leads to bulk magnetic asymmetry and gradient Dzyaloshinskii–Moriya interaction (g-DMI). [24,25,30]

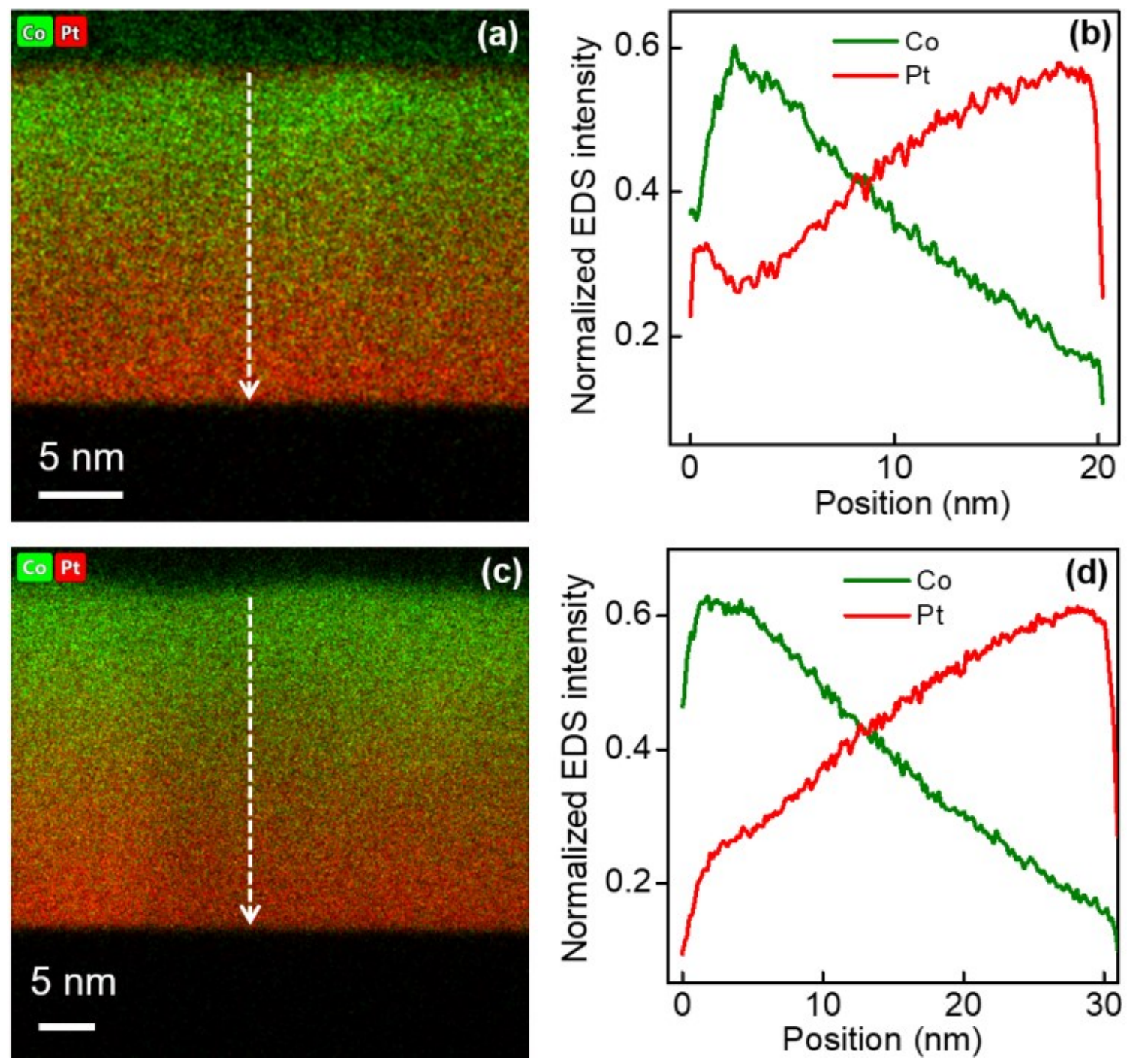

**Figure S1.2:** STEM-EDS image of $\Delta x$ = +50% 20 nm (a) and 30 nm (c) thick g-CoPt films with plotted contribution of Co (green) and Pt (red). Normalized EDS intensity for Pt (red solid line) and Co (green solid line) for $\Delta x$ = +50% 20 nm (b) and 30 nm (d) thick g-CoPt films, measured over the entire area along the arrows in (a) and (c), respectively.

## S2. Representative Brillouin light scattering (BLS) measurements of 30 nm g-CoPt films

Due to the relationship between net g-DMI and effective gradient $\Delta x/t$, the non-reciprocal frequency shift $\Delta f$ between the stokes and anti-stokes peak becomes almost indistinguishable in the measurement for $\Delta x$ = +50% 30 nm g-CoPt film (see Figure S2.1a). For $\Delta x$ = -50% 30 nm g-CoPt film (Figure S2.1b), the BLS intensity is very weak to extract $\Delta f$, that may be explained by the multidomain state reduces the ordered propagation of magnons.

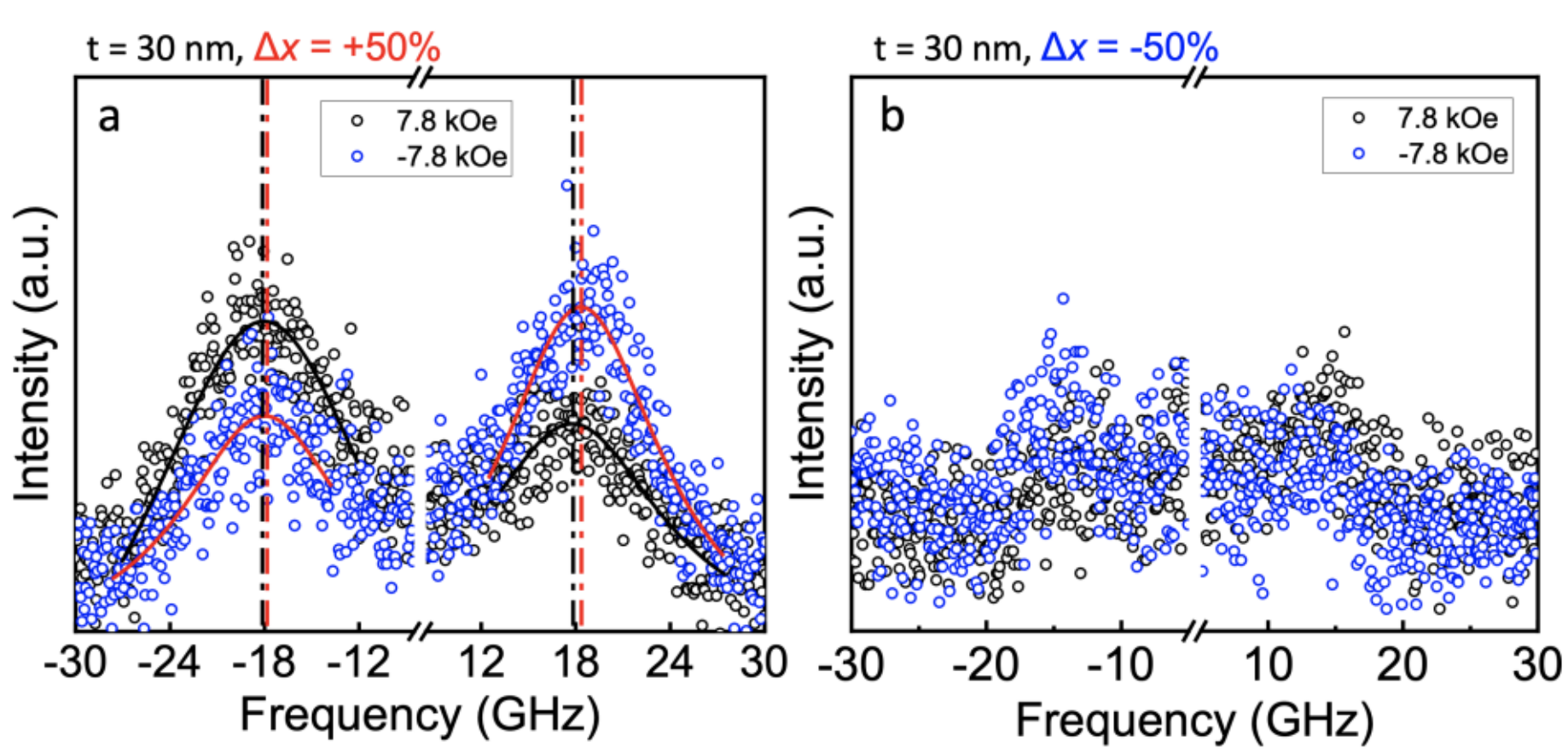

**Figure S2.1:** BLS spectra of 30 nm, $\Delta x$ = +50% (a) and $\Delta x$ = -50% (b) g-CoPt films. Due to the low signal to noise ratio, an estimation of DMI could not be obtained for $\Delta x$ = -50% g-CoPt.

**S3. MFM imaging and $\Delta\rho_{xy}(H)$ curves in g-CoPt thin films with $\Delta x$ = -50%**

To measure $\Delta\rho_{xy}(H)$ correctly, we calibrated both Physical Property Measurement System (PPMS) and magneto-optical Kerr effect (MOKE) setups using a reference FePt single layer with a strong perpendicular magnetic anisotropy (PMA). The coercive fields of FePt should be identical when measured in both systems, thus enabling us to calibrate the magnetic field scales accordingly. Figures S3.1a and S3.1b present the repeatedly measured MOKE hysteresis loops and Hall resistance curves at different regions of the sample, respectively, with sweep rates of 25 Oe/s for MOKE and 20 Oe/s for Hall measurements. We also varied the field sweep direction during the measurements. To account for differences in field strength between the PPMS (used for transport measurement) and the MOKE system, we compared the coercive fields extracted from the averaged MOKE and Hall curves. A consistent scaling relation was observed: $H_{\mathrm{MOKE}} \times 0.85 = H_{\mathrm{PPMS}}$. Based on this, we applied a field correction factor to analyze the residual Hall resistivity $\Delta\rho_{xy}$.

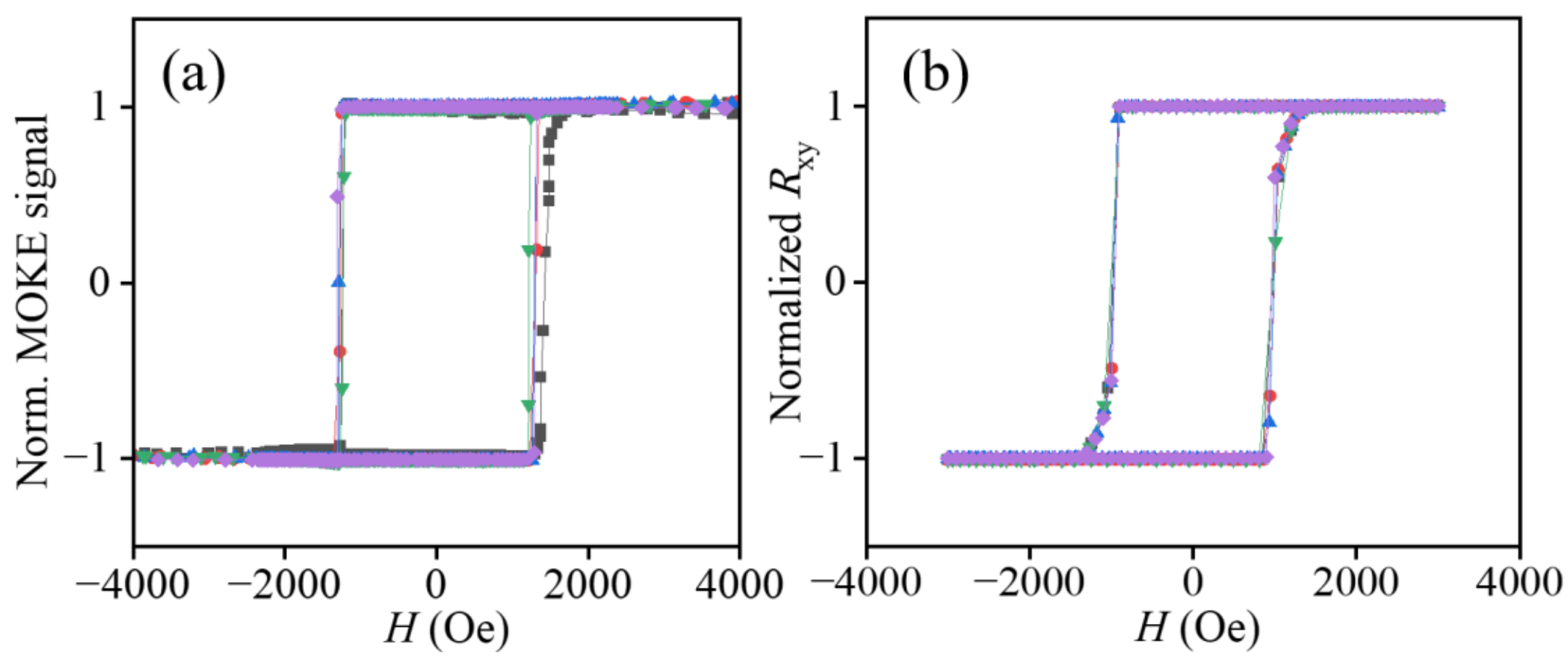

**Figure S3.1:** (a) MOKE hysteresis loops and (b) Hall resistance curves taken at different regions of the FePt film (scattered curves).

Characterization by $\Delta\rho_{xy}(H)$ curves included topological hall effect (THE) and magnetic force microscopy (MFM) measurements were also conducted on $\Delta x$ = -50% 20 nm (Figure S3.2a) and 30 nm (Figure S3.2b) g-CoPt/STO films. The size and density of the spin textures are discussed in the main text. The density of skyrmions in $\Delta x$ = -50% 30 nm g-CoPt film is very low (~0.4 $\mu m^{-2}$) in comparison to +50% 30 nm g-CoPt (~5 $\mu m^{-2}$). This correlates well with BLS measurements, where g-DMI cannot be extracted in $\Delta x$ = -50% films due to the weak magnetic anisotropy (discussed above). However, the $\Delta\rho_{xy}(H)$ amplitude of both $\Delta x$ = ±50% films is comparable, confirming the challenges of using only $\Delta\rho_{xy}(H)$ curves to extract information about chiral spin textures.[39]

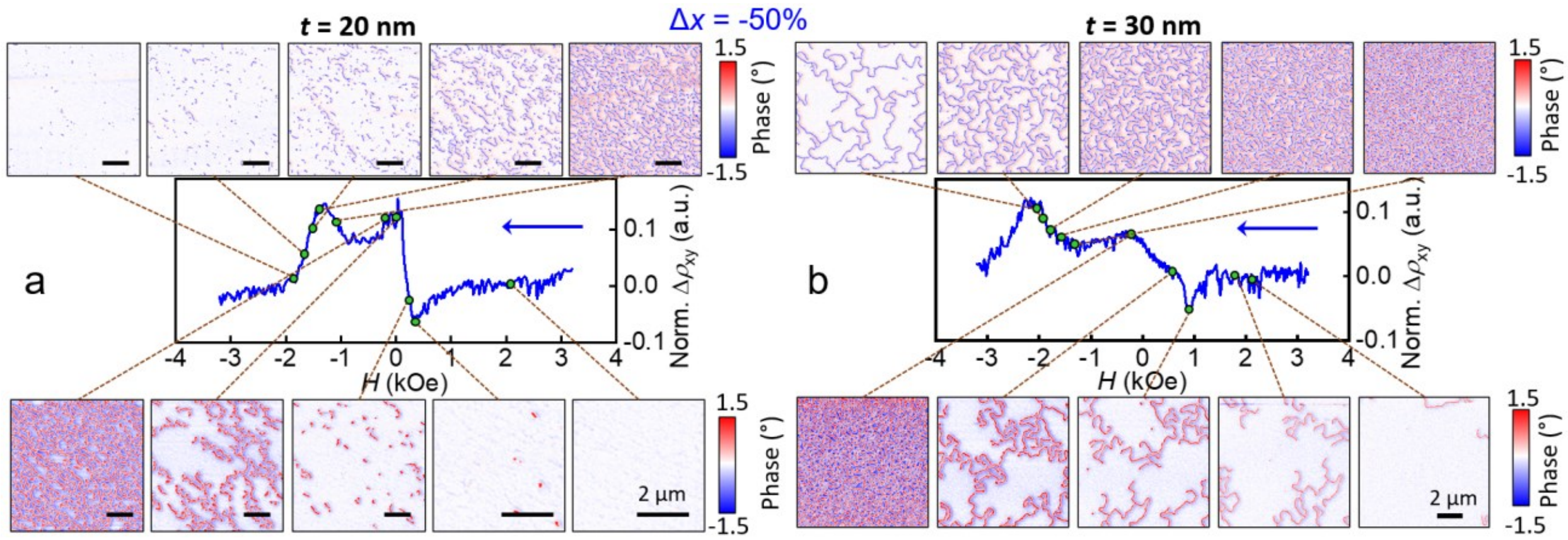

**Figure S.3.2:** Correlative THE and MFM of 20 nm (a) and 30 nm (b) $\Delta x$ = -50% g-CoPt films. Blue arrows represent the field sweep directions of both THE and MFM. The scale bars in all MFM maps in (a) is 2 μm.

MFM measurements conducted on $\Delta x$ = -50% 30 nm g-CoPt film grown on sapphire (Figure S3.3) confirm the absence of magnetic skyrmions that may be explained by the effect of substrate (lattice mismatch) in particular for the $Co_3Pt \rightarrow CoPt_3$.[30]

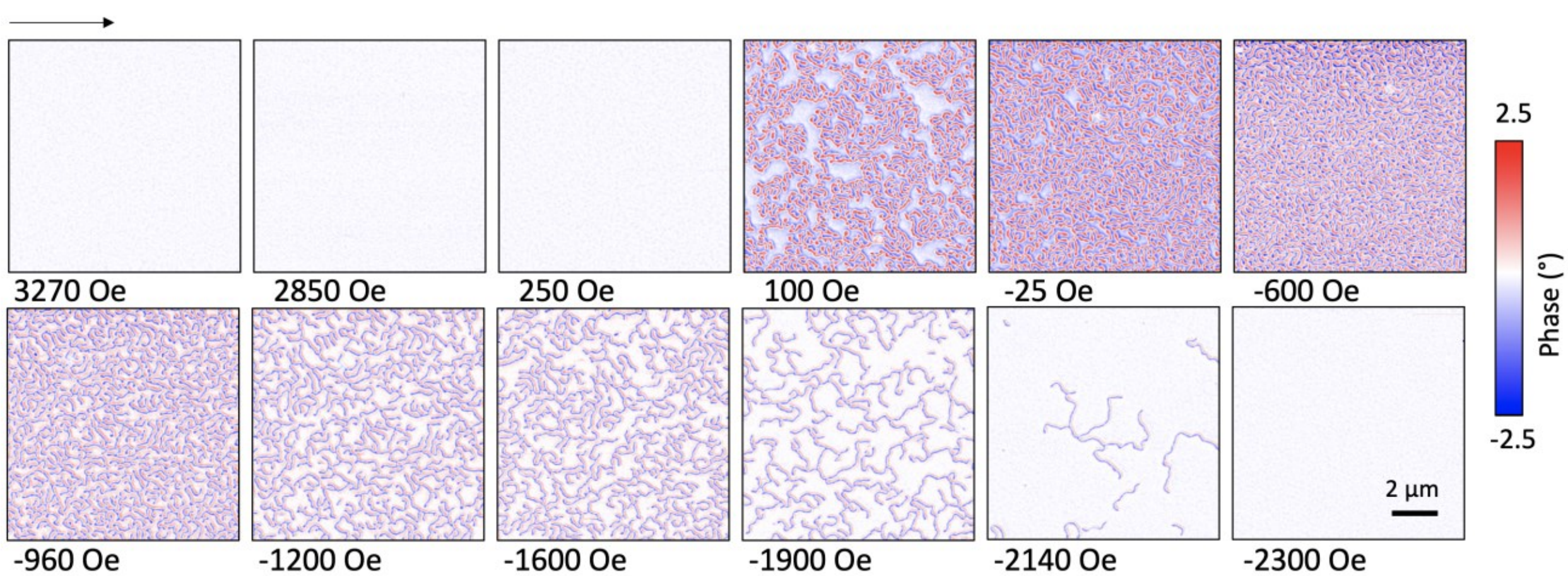

**Figure S3.3:** MFM images of 30 nm, $\Delta x$ = -50% g-CoPt grown on $Al_2O_3$ substrate. The black arrow represents the field sweep direction of MFM measurements

## S4. Additional Micromagnetic Simulations

To investigate the stability of skyrmions under various applied magnetic fields and g-DMI strengths, we simulated phase diagrams showing the skyrmion radius and average domain wall (DW) angle, as presented in Figure S4.1, where isolated skyrmions in our simulations could be well represented as 360 degrees DWs. As expected from skyrmion energetics, an applied magnetic field (parallel to the ferromagnetic background magnetization) decreases the skyrmion radius. Above a certain field strength, the skyrmion state collapses (i.e., the radius vanishes), as observed experimentally from nitrogen vacancy (NV) measurements in 10 nm g-CoPt films.[30] Conversely, if the applied magnetic field is too low, maze states are energetically preferred over isolated skyrmions (indicated by the dark red regions in Figures S4.1a and S4.1c).

Figures S4.1b and S4.1d illustrate that increasing the g-DMI strength decreases the average DW angle, bringing it closer to the Néel type. This indicates a transition region from Bloch-type towards Néel-type character, corresponding to hybrid skyrmions.[51] Furthermore, as the applied magnetic field *H* increases, the skyrmion shrinks, thereby reducing dipolar energy contribution relative to the DMI energy. This effect also contributes to a decrease in the DW angle (a shift towards Néel type). This behavior arises because the dipolar energy contribution favors Bloch-type DWs, while the DMI contribution favors Néel-type DWs.[52] The competition between these energy terms determines the equilibrium DW angle, often resulting in a hybrid skyrmion state in our simulations.

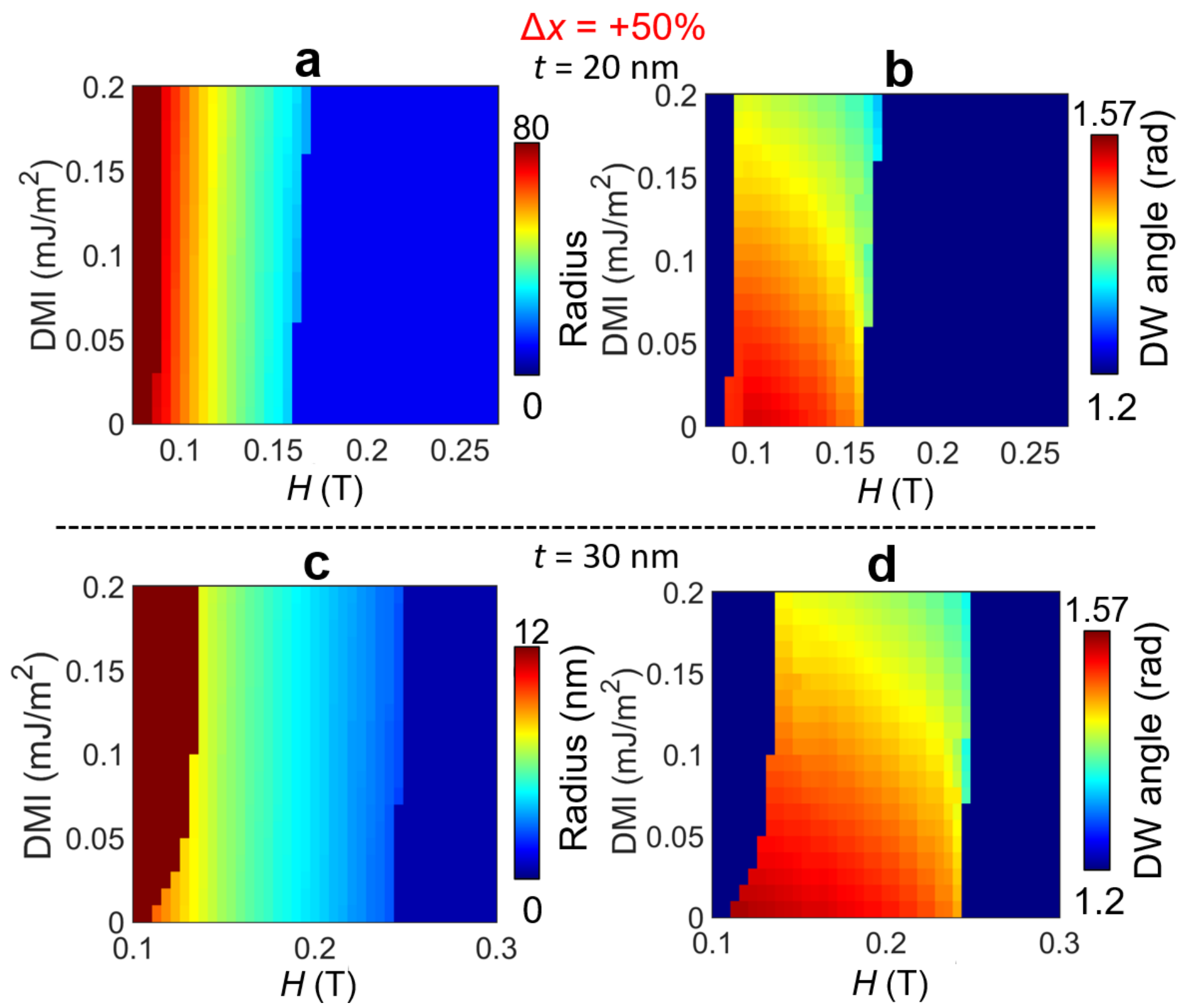

**Figure S4.1:** Systematically simulated DMI vs H phase diagrams. Calculated radius (a) and domain wall angle (b) of $\Delta x$ = +50% 20 nm g-CoPt film ($K_u$ =440 kJ/m$^3$) as a function of g-DMI strength and applied magnetic field amplitude. Calculated radius (c) and domain wall angle (d) of $\Delta x$ = +50% 30 nm g-CoPt films ($K_u$ =480 kJ/m$^3$) as a function of g-DMI strength and applied magnetic field amplitude.

In Figure S4.2, we examine the dependence of the skyrmion radius on the applied magnetic field *H* for different effective magnetic anisotropy density ($K_u$) values. Increasing $K_u$ lowers the minimum applied magnetic field required to stabilize the skyrmion state relative to the maze state. Conversely, the collapse field (the field above which the skyrmion is no longer stable, i.e., the radius approaches zero) increases as $K_u$ decreases. The simulation dimensions are 1.2 µm × 1.2 µm × 20 nm (30 nm).

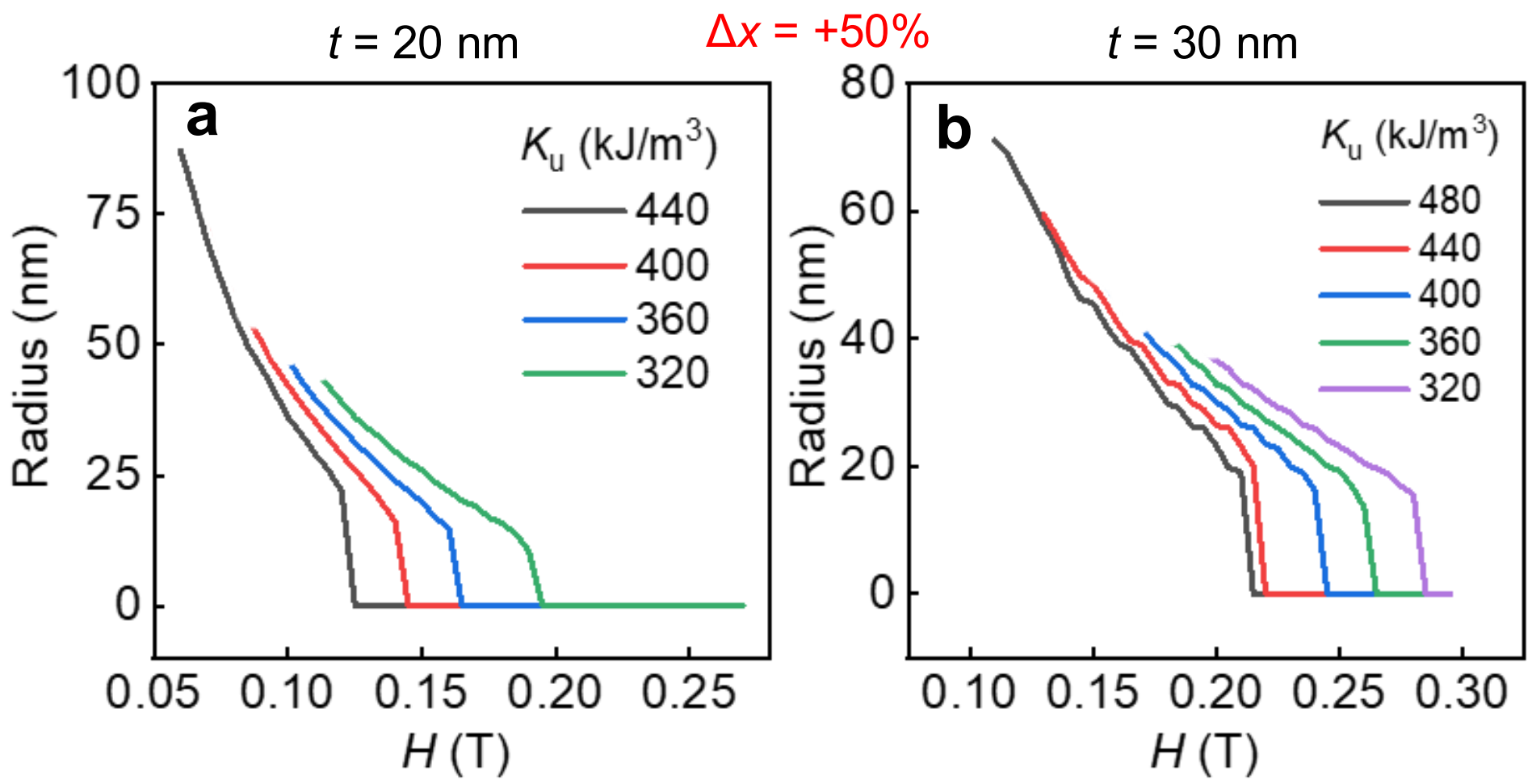

**Figure S4.2.** Skyrmion radius vs applied magnetic field for varying effective anisotropy in 20 nm (a) and 30 nm (b) $\Delta x$ = +50% g-CoPt 20 nm correspond to $K_u$ =440 kJ/m$^3$ and $\Delta x$ = +50% g-CoPt 30 nm correspond to $K_u$ = 480 kJ/m$^3$.

Next, we use micromagnetic simulations to assess the range of magnetic fields in which isolated skyrmions can be observed, establishing the low-field bound. Figure S4.3 shows the area density of the energy difference between maze state and uniform state over the simulation system size (1.2 μm × 1.2 μm). The negative values of the energy difference at small magnetic fields indicate that the maze state is more stable than the uniform state. As a result, in this range of magnetic fields, one cannot consider isolated skyrmions in the traditional sense, as they typically arise as metastable states on top of the uniform state. Nevertheless, in this region of magnetic fields one might expect coexistence of skyrmions and maze domains. This behavior agrees with some of our experimental findings where for some applied fields both skyrmion and maze domains coexist in the system, see, e.g., Figure S3.1.

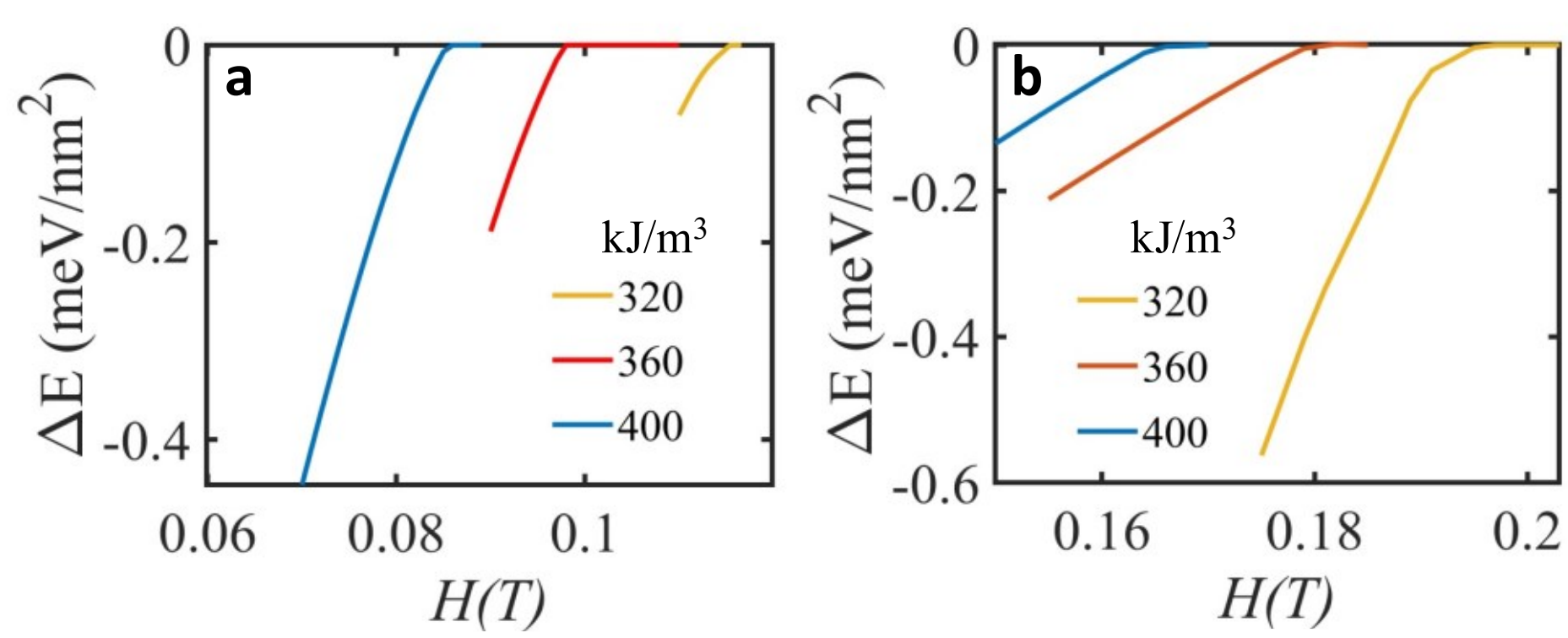

**Figure S4.3.** The energy difference between the maze and uniform states for 20 nm (a) and 30 nm (b).

Other than the skyrmions and maze domains, trivial bubbles can also be stabilized in our systems. Figure S4.4a shows stabilization of the trivial bubble state at same magnetic fields as skyrmions for $\Delta x$ = +50% g-CoPt 20 nm. As the magnetic field increases, bubbles disappear at lower field strengths than skyrmions do, indicating reduced stability for bubbles and favoring the formation of skyrmion states. Figure S4.4b displays the energies of the skyrmion and the bubble relative to the ferromagnetic state (energy set to zero). Consistent with expectations from Figure S4.4a, trivial bubbles exhibit a higher energy, implying a lower energy barrier compared to skyrmions.

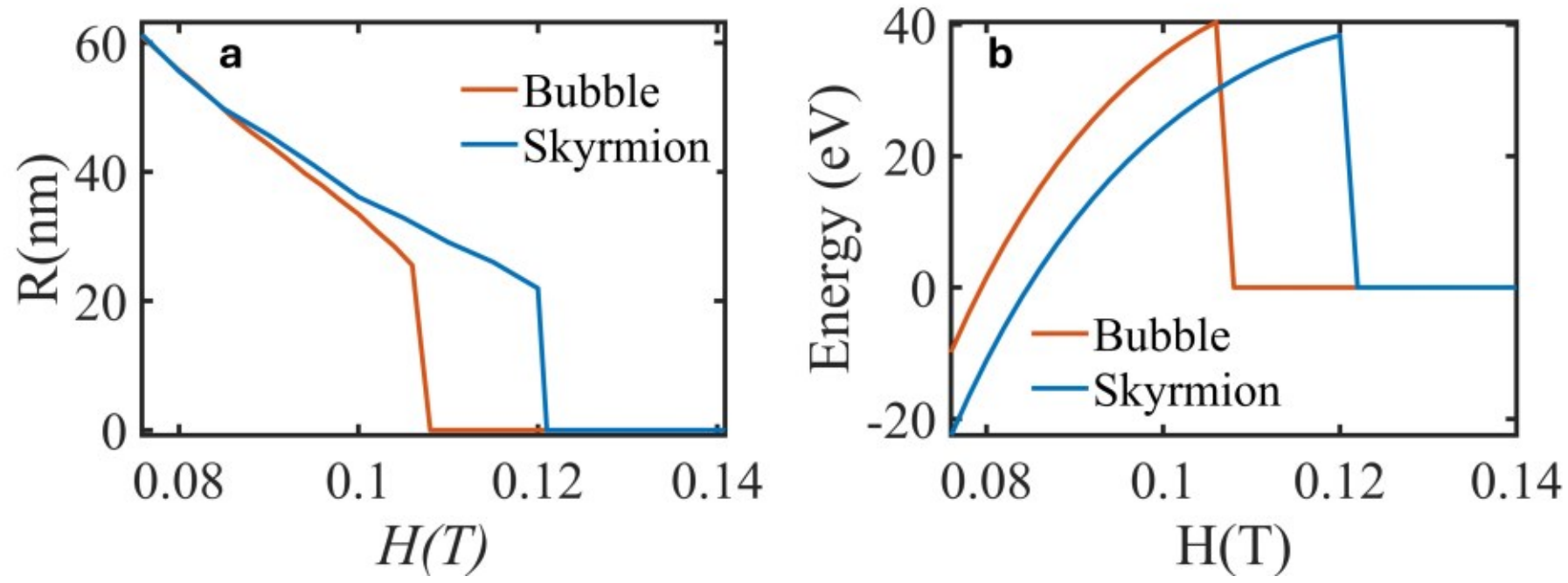

**Figure S4.4.** The radius of trivial bubble and skyrmion for $\Delta x$ = +50% g-CoPt 20 nm ($K_u$ =440 kJ/m$^3$) as applied field is increased (a) and energy of bubble and skyrmion shown in (b).

## S5. Minimum Energy Path Calculations

In ferromagnetic films dipolar interactions may stabilize trivial magnetic bubbles or other textures such as antiskyrmions,[46] in addition to skyrmions. This makes it difficult to determine whether the solitons observed in the experiment are in fact skyrmions. To help support our conclusions we have carried out minimum energy path (MEP) calculations by the string method, using modified Mumax3 to examine the annihilation processes of skyrmions and trivial bubbles.[44] These calculations were used to find the least energy path for the decay of skyrmion and bubble initial states, as well as the energy barrier of the decay process. Since the rate of decay is given by $\Gamma = f e^{-\beta \Delta E}$ where $\Delta E$ is the energy barrier and $f$ is the attempt frequency, this may provide some insight into the relative populations of these two types of solitons.[53]

The calculations were performed by relaxing an initial state of a skyrmion by the steepest conjugate gradient method, followed by running the MEP solver over 800 cycles, with 32 images in the transition chain. These calculations show that a skyrmion decaying to a uniform magnetic state could transition through a metastable trivial bubble state, before decaying to the ground state. Further, we see that the energy barrier for the transition from a trivial bubble to a skyrmion is much smaller than the reverse process. This suggests that while trivial bubbles may account for some fraction of the observed solitons in the sample, we can expect skyrmions to account for the majority.

Figure S5.1 shows the calculated energy and skyrmion winding number $N_{sk}$ of the system as a function of the image number during the skyrmion decay for the $\Delta x$ = +50% 20 nm g-CoPt film at a magnetic field of 85 mT. Here the initial state was chosen to be a skyrmion, while the final state is the homogeneous ground state. We see three energy minima in the plot, in order from left to right: trivial bubble, skyrmion, and homogeneous. The transition from the skyrmion to the bubble during the decay process does not require any initial input to the simulation; it occurs naturally during the decay of the skyrmion along the minimum energy path. The energy barriers for these parameters are much larger than room temperature, which suggests that bubbles and skyrmions are likely to coexist within the range of magnetic field where bubbles are metastable, as shown in figure S4.4. Given the magnitude of the energy difference between a skyrmion and a bubble, as compared to the energy scale determined by temperature, we conclude that the appearance of bubbles is suppressed and may further lead to their decay into skyrmions.

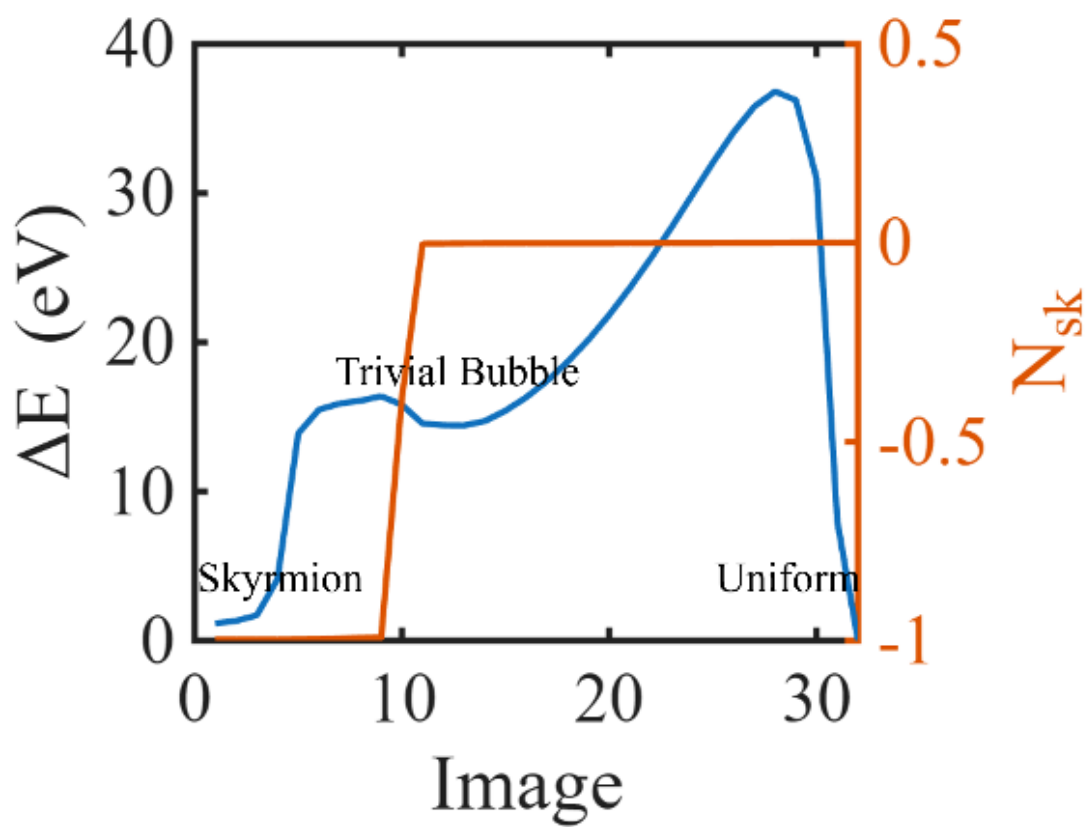

**Figure S5.1.** Energy during the transition from trivial bubble to uniform along the least energy path. Material parameters correspond to the 20 nm +50% gradient film, with an out of plane magnetic field of 85 mT. The lattice dimensions are 300 × 300 × 8 with a lattice constant of 2.5 nm. $K_u = 440\ kJ\ m^{-3}$, $D = 0.12\ mJ\ m^{-2}$. $N_{sk}$ is the skyrmion winding number. The $x$-axis corresponds to the image number in the transition chain and can be interpreted as the reaction coordinate of the process.

We also compare the energy barrier for annihilation of a skyrmion for different magnetic fields, to the energy difference between the homogeneous state and skyrmion, shown in Figure S5.2. We can expect that when the energy difference between the skyrmion and the homogeneous state is positive, and the barrier for the transition is nonzero that skyrmions will be metastable states, with finite lifetimes. When the energy difference between the skyrmion and the homogeneous state is negative, it shows that homogeneous magnetization is no longer the ground state, and that the true ground state should be either maze domains or a skyrmion lattice. The point where the energy difference becomes negative should correspond roughly to the point where the skyrmion becomes unstable and is likely to expand into a maze.[54] Therefore, we should only expect to observe skyrmions within a finite range of magnetic fields, where the energy difference is positive, and the energy barrier is greater than zero.

Our calculations show qualitative agreement with the measurements presented in Figure 5b (main text), where skyrmions in the 20 nm positive gradient sample are only observed within the range of 150 to 210 mT (or 1.5 to 3 kOe). The calculations underestimate this range slightly; however this may be due to the lack of disorder, e.g., originating from nonuniformity of the sample. The inclusion of disorder would result in pinning which prevents the expansion of skyrmions at lower fields, and the decay of skyrmions at higher fields, thus affecting their stability.[47]

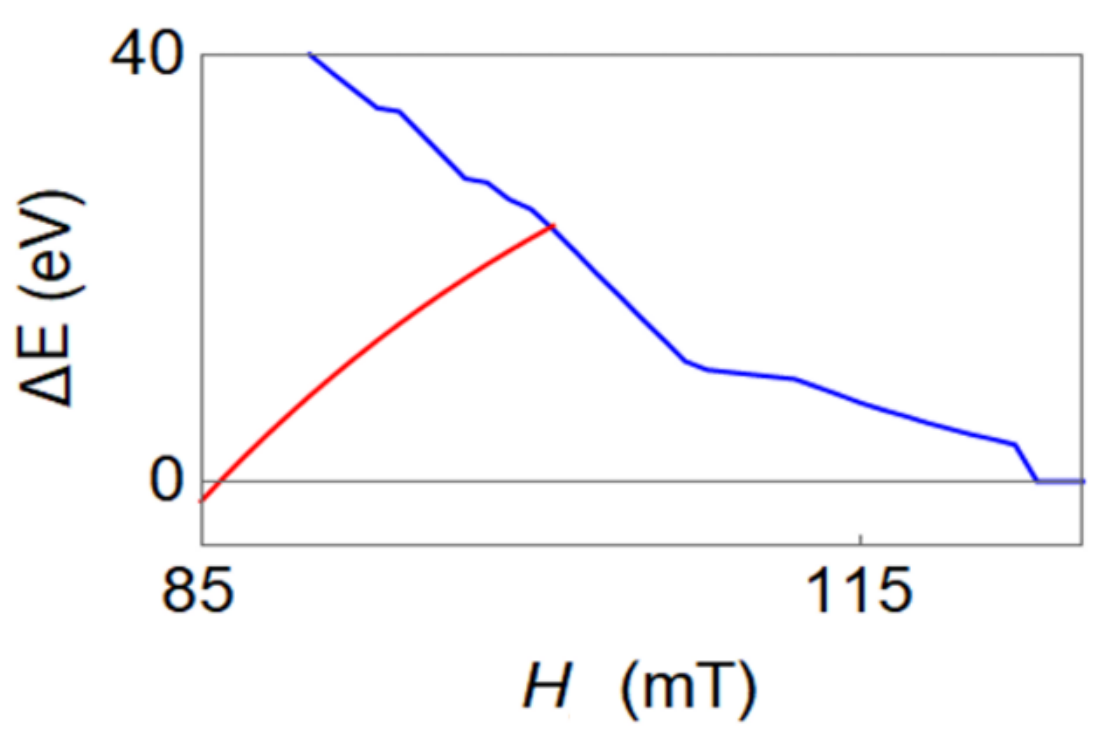

**Figure S5.2.** Energy barrier for transition from skyrmion to homogeneous (blue) compared to energy difference between skyrmion and homogeneous (red) for $\Delta x = +50\%$ 20 nm g-CoPt single layer. Results are obtained after 200 cycles with 30 images in the chain. $K_u = 440\ \text{kJ/m}^3$, $D = 0.12\ \text{mJ/m}^2$.

Figure S5.3 shows the simulated decay of skyrmions to homogeneous magnetization, for $\Delta x = +50$ 20 nm g-CoPt film for several image snapshots. Skyrmion initially transitions to trivial bubble state, then the bubble gradually decreases in size until the uniform state is reached.

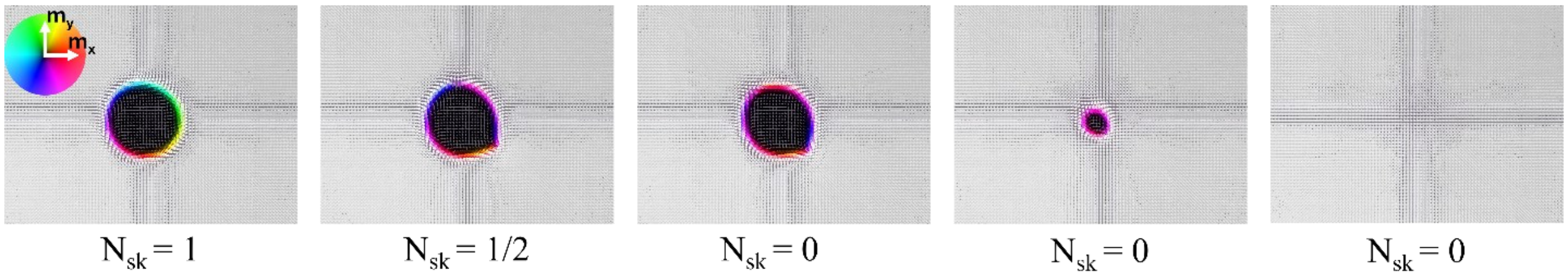

**Figure S5.3.** Images of a skyrmion to trivial bubble transmission and subsequent decay to uniform state done at *H* amplitude of 85 mT in +50% 20 nm g-CoPt film. White (Black) represents positive (negative) out-of-plane magnetization.

## References

[1] A. Fert, V. Cros, J. Sampaio, *Nature Nanotech* **2013**, *8*, 152.
[2] A. Fert, N. Reyren, V. Cros, *Nat Rev Mater* **2017**, *2*, 1.
[3] T. Moriya, *Phys. Rev.* **1960**, *120*, 91.
[4] I. Dzyaloshinsky, *Journal of Physics and Chemistry of Solids* **1958**, *4*, 241.
[5] S. Heinze, K. Von Bergmann, M. Menzel, J. Brede, A. Kubetzka, R. Wiesendanger, G. Bihlmayer, S. Blügel, *Nature Phys* **2011**, *7*, 713.
[6] N. Romming, C. Hanneken, M. Menzel, J. E. Bickel, B. Wolter, K. Von Bergmann, A. Kubetzka, R. Wiesendanger, *Science* **2013**, *341*, 636.
[7] C. Moreau-Luchaire, C. Moutafis, N. Reyren, J. Sampaio, C. A. F. Vaz, N. Van Horne, K. Bouzehouane, K. Garcia, C. Deranlot, P. Warnicke, P. Wohlhüter, J.-M. George, M. Weigand, J. Raabe, V. Cros, A. Fert, *Nature Nanotech* **2016**, *11*, 444.
[8] S. Woo, K. Litzius, B. Krüger, M.-Y. Im, L. Caretta, K. Richter, M. Mann, A. Krone, R. M. Reeve, M. Weigand, P. Agrawal, I. Lemesh, M.-A. Mawass, P. Fischer, M. Kläui, G. S. D. Beach, *Nature Mater* **2016**, *15*, 501.
[9] O. Boulle, J. Vogel, H. Yang, S. Pizzini, D. De Souza Chaves, A. Locatelli, T. O. Menteş, A. Sala, L. D. Buda-Prejbeanu, O. Klein, M. Belmeguenai, Y. Roussigné, A. Stashkevich, S. M. Chérif, L. Aballe, M. Foerster, M. Chshiev, S. Auffret, I. M. Miron, G. Gaudin, *Nature Nanotech* **2016**, *11*, 449.
[10] A. Soumyanarayanan, M. Raju, A. L. Gonzalez Oyarce, A. K. C. Tan, M.-Y. Im, A. P. Petrović, P. Ho, K. H. Khoo, M. Tran, C. K. Gan, F. Ernult, C. Panagopoulos, *Nature Mater* **2017**, *16*, 898.
[11] J. Park, T. Kim, G. W. Kim, V. Bessonov, A. Telegin, I. G. Iliushin, A. A. Pervishko, D. Yudin, A. Yu. Samardak, A. V. Ognev, A. S. Samardak, J. Cho, Y. K. Kim, *Acta Materialia* **2022**, *241*, 118383.
[12] S. Mühlbauer, B. Binz, F. Jonietz, C. Pfleiderer, A. Rosch, A. Neubauer, R. Georgii, P. Böni, *Science* **2009**, *323*, 915.
[13] Y. Li, N. Kanazawa, X. Z. Yu, A. Tsukazaki, M. Kawasaki, M. Ichikawa, X. F. Jin, F. Kagawa, Y. Tokura, *Phys. Rev. Lett.* **2013**, *110*, 117202.
[14] X. Z. Yu, Y. Onose, N. Kanazawa, J. H. Park, J. H. Han, Y. Matsui, N. Nagaosa, Y. Tokura, *Nature* **2010**, *465*, 901.
[15] X. Z. Yu, N. Kanazawa, Y. Onose, K. Kimoto, W. Z. Zhang, S. Ishiwata, Y. Matsui, Y. Tokura, *Nature Mater* **2011**, *10*, 106.
[16] F. Zheng, N. S. Kiselev, L. Yang, V. M. Kuchkin, F. N. Rybakov, S. Blügel, R. E. Dunin-Borkowski, *Nat. Phys.* **2022**, *18*, 863.

[17] T. Srivastava, M. Schott, R. Juge, V. Křižáková, M. Belmeguenai, Y. Roussigné, A. Bernand-Mantel, L. Ranno, S. Pizzini, S.-M. Chérif, A. Stashkevich, S. Auffret, O. Boulle, G. Gaudin, M. Chshiev, C. Baraduc, H. Béa, *Nano Lett.* **2018**, *18*, 4871.
[18] S. Li, X. Lin, P. Li, S. Zhao, Z. Si, G. Wei, B. Koopmans, R. Lavrijsen, W. Zhao, *ACS Appl. Mater. Interfaces* **2023**, *15*, 39946.
[19] N. S. Gusev, A. V. Sadovnikov, S. A. Nikitov, M. V. Sapozhnikov, O. G. Udalov, *Phys. Rev. Lett.* **2020**, *124*, 157202.
[20] S. Li, S. Zhao, D. M. F. Hartmann, W. Yang, X. Lin, W. Zhao, R. Lavrijsen, *Applied Physics Letters* **2025**, *126*, 162401.
[21] N. Kato, M. Kawaguchi, Y.-C. Lau, T. Kikuchi, Y. Nakatani, M. Hayashi, *Phys. Rev. Lett.* **2019**, *122*, 257205.
[22] S. Kim, K. Ueda, G. Go, P.-H. Jang, K.-J. Lee, A. Belabbes, A. Manchon, M. Suzuki, Y. Kotani, T. Nakamura, K. Nakamura, T. Koyama, D. Chiba, K. T. Yamada, D.-H. Kim, T. Moriyama, K.-J. Kim, T. Ono, *Nat Commun* **2018**, *9*, 1648.
[23] S. Zhang, J. Zhang, Y. Wen, E. M. Chudnovsky, X. Zhang, *Commun Phys* **2018**, *1*, 36.
[24] J. Liang, M. Chshiev, A. Fert, H. Yang, *Nano Lett.* **2022**, *22*, 10128.
[25] Q. Zhang, J. Liang, K. Bi, L. Zhao, H. Bai, Q. Cui, H.-A. Zhou, H. Bai, H. Feng, W. Song, G. Chai, O. Gladii, H. Schultheiss, T. Zhu, J. Zhang, Y. Peng, H. Yang, W. Jiang, *Phys. Rev. Lett.* **2022**, *128*, 167202.
[26] A. Laraoui, K. Ambal, *Applied Physics Letters* **2022**, *121*, 060502.
[27] A. Erickson, S. Q. A. Shah, A. Mahmood, I. Fescenko, R. Timalsina, C. Binek, A. Laraoui, *RSC Adv.* **2022**, *13*, 178.
[28] A. Erickson, S. Q. A. Shah, A. Mahmood, P. Buragohain, I. Fescenko, A. Gruverman, C. Binek, A. Laraoui, *Adv Funct Materials* **2024**, 2408542.
[29] S.-H. Liou, in *Handbook of Advanced Magnetic Materials* (Eds.: Y. Liu, D. J. Sellmyer, D. Shindo), Springer US, Boston, MA, **2006**, pp. 374–396.
[30] A. Erickson, Q. Zhang, H. Vakili, C. Li, S. Sarin, S. Lamichhane, L. Jia, I. Fescenko, E. Schwartz, S.-H. Liou, J. E. Shield, G. Chai, A. A. Kovalev, J. Chen, A. Laraoui, *ACS Nano* **2024**, *18*, 31261.
[31] Z. Zheng, Y. Zhang, V. Lopez-Dominguez, L. Sánchez-Tejerina, J. Shi, X. Feng, L. Chen, Z. Wang, Z. Zhang, K. Zhang, B. Hong, Y. Xu, Y. Zhang, M. Carpentieri, A. Fert, G. Finocchio, W. Zhao, P. Khalili Amiri, *Nat Commun* **2021**, *12*, 4555.
[32] O. Kapusta, A. Zelenakova, J. Bednarcik, V. Zelenak, Slovak University Of Technology, Slovakia, **2017**, p. 346.
[33] M. Ohtake, S. Ouchi, F. Kirino, M. Futamoto, *IEEE Transactions on Magnetics* **2012**, *48*, 3595.
[34] M. T. Johnson, P. J. H. Bloemen, F. J. A. D. Broeder, J. J. D. Vries, *Rep. Prog. Phys.* **1996**, *59*, 1409.
[35] X. Lu, D. E. Laughlin, J.-G. (Jimmy) Zhu, *Journal of Magnetism and Magnetic Materials* **2019**, *491*, 165570.
[36] M. Ohtake, D. Suzuki, M. Futamoto, *Journal of Applied Physics* **2014**, *115*, 17C116.
[37] Y. Yang, B. Varghese, H. K. Tan, S. K. Wong, S. N. Piramanayagam, *Journal of Applied Physics* **2014**, *115*, 083910.
[38] M. Raju, A. Yagil, A. Soumyanarayanan, A. K. C. Tan, A. Almoalem, F. Ma, O. M. Auslaender, C. Panagopoulos, *Nat Commun* **2019**, *10*, 696.
[39] G. Kimbell, C. Kim, W. Wu, M. Cuoco, J. W. A. Robinson, *Commun Mater* **2022**, *3*, 19.

[40] L. Wang, Q. Feng, Y. Kim, R. Kim, K. H. Lee, S. D. Pollard, Y. J. Shin, H. Zhou, W. Peng, D. Lee, W. Meng, H. Yang, J. H. Han, M. Kim, Q. Lu, T. W. Noh, *Nature Mater* **2018**, *17*, 1087.
[41] A. Rajanikanth, S. Kasai, N. Ohshima, K. Hono, *Applied Physics Letters* **2010**, *97*, 022505.
[42] F. Büttner, I. Lemesh, G. S. D. Beach, *Sci Rep* **2018**, *8*, 4464.
[43] A. Vansteenkiste, J. Leliaert, M. Dvornik, M. Helsen, F. Garcia-Sanchez, B. Van Waeyenberge, *AIP Advances* **2014**, *4*, 107133.
[44] W. E, W. Ren, E. Vanden-Eijnden, *The Journal of Chemical Physics* **2007**, *126*, 164103.
[45] U. Güngördü, R. Nepal, O. A. Tretiakov, K. Belashchenko, A. A. Kovalev, *Phys. Rev. B* **2016**, *93*, 064428.
[46] A. A. Kovalev, S. Sandhoefner, *Front. Phys.* **2018**, *6*, 98.
[47] R. Gruber, J. Zázvorka, M. A. Brems, D. R. Rodrigues, T. Dohi, N. Kerber, B. Seng, M. Vafaee, K. Everschor-Sitte, P. Virnau, M. Kläui, *Nat Commun* **2022**, *13*, 3144.
[48] Y. Dovzhenko, F. Casola, S. Schlotter, T. X. Zhou, F. Büttner, R. L. Walsworth, G. S. D. Beach, A. Yacoby, *Nat Commun* **2018**, *9*, 2712.
[49] G. Chen, *Nature Phys* **2017**, *13*, 112.
[50] F. Ajejas, Y. Sassi, W. Legrand, S. Collin, J. Peña Garcia, A. Thiaville, S. Pizzini, N. Reyren, V. Cros, A. Fert, *Phys. Rev. Materials* **2022**, *6*, L071401.
[51] H. Vakili, Y. Xie, A. W. Ghosh, *Phys. Rev. B* **2020**, *102*, 174420.
[52] C.-E. Fillion, J. Fischer, R. Kumar, A. Fassatoui, S. Pizzini, L. Ranno, D. Ourdani, M. Belmeguenai, Y. Roussigné, S.-M. Chérif, S. Auffret, I. Joumard, O. Boulle, G. Gaudin, L. Buda-Prejbeanu, C. Baraduc, H. Béa, *Nat Commun* **2022**, *13*, 5257.
[53] P. F. Bessarab, V. M. Uzdin, H. Jónsson, *Phys. Rev. B* **2012**, *85*, 184409.
[54] H. T. Diep, S. El Hog, A. Bailly-Reyre, *AIP Advances* **2018**, *8*, 055707.